\documentclass {article}

\usepackage{alltt,multirow,booktabs,float}  %
\usepackage{amssymb,amsfonts,graphicx,color,setspace}
\usepackage{fullpage}

\newcommand{\qed}{\nobreak \ifvmode \relax \else
      \ifdim\lastskip<1.5em \hskip-\lastskip
      \hskip1.5em plus0em minus0.5em \fi \nobreak
      $\Box$\fi}

\begin{document}
\title{Small-Sample Behavior of Novel Phase~I Cancer Trial Designs}
\author{Assaf P. Oron and Peter D. Hoff \\ Modified Version of Clinical Trials Journal Submission \\
Correspondence: Assaf P. Oron, Seattle Children's Research Institute \\ 2001 Eighth Avenue, Suite 500, Seattle, WA 98121, U.S.A. \\
\emph{assaf.oron@seattlechildrens.org} \\ }

\maketitle
\begin{abstract}
\onehalfspacing
\textbf{Background:} Novel dose-finding designs for Phase~I cancer clinical trials, using estimation to assign the best estimated maximum-tolerated-dose (MTD) at each point in the experiment, most prominently via Bayesian techniques, have been widely discussed and promoted since 1990.

\textbf{Purpose:} To examine the small-sample behavior of these ``Bayesian Phase~I'' (BP1) designs, and also of non-Bayesian designs sharing the same main ``long-memory'' traits of using likelihood estimation and assigning the estimated MTD to the next patient. We refer to the larger family that includes both BP1 and similar non-Bayesian designs, as LMP1 (``long-memory Phase~I'').

\textbf{Methods:} Data from several recently published experiments are presented and discussed, and LMP1's operating principles are explained. Simulation studies compare the small-sample behavior of LMP1s with short-memory ``up-and-down'' (U\&D) designs.

\textbf{Results:} LMP1s and U\&D achieved similar success rates in finding the MTD. However, for all LMP1s examined, the number $n^*$ of cohorts treated at the true MTD was highly variable between simulated experiments drawn from the same toxicity-threshold distribution, especially when compared with U\&D. Further investigation using the same set of thresholds in permuted order, suggests that this LMP1 behavior is driven by a strong sensitivity to the order in which participants enter the experiment. This sensitivity is related to LMP1's tendency to settle early on a specific dose level, a tendency caused by their ``winner-takes-all'' dose assignment rule, which grants the early cohorts a disproportionately large influence. Additionally for the Bayesian BP1 family, the manner in which they use model and prior specification to engineer early-cohort behavior generates a system of interdependencies that is hard to control, and in certain ways contradicts the rationale of Bayesian modeling.

\textbf{Limitations:} While the numerical evidence for LMP1's high run-to-run variability is broad, and sensible explanations for it are provided, we do not present a theoretical proof of the phenomenon.

\textbf{Conclusions:} Method developers, analysts and practitioners should be aware of LMP1's variability and order-sensitivity, and of the factors driving them. In particular, they should be informed that settling on a single dose does not guarantee that this dose is the MTD. Presently, U\&D designs offer a simpler and more stable alternative for the sample sizes of $10-40$ patients used in most Phase~I trials. We also suggest that the field's paradigm change from dose-selection to dose-estimation.
\end{abstract}


\doublespacing

\section{Introduction}\label{sec:crm1}

Over the past two decades, numerous novel dose-finding designs employing Bayesian calculations, such as continual reassessment method \cite{OQuigleyEtAl90} and escalation with overdose control \cite{BabbEtAl98}, have been developed for Phase~I cancer trials. The hallmark of these designs is estimation of the dose-toxicity function after each cohort, in order to assign the estimated Maximum Tolerated Dose (MTD) to the next cohort. These ``Bayesian Phase~I'' designs (in short: BP1s) have been joined by novel non-Bayesian designs employing this principle \cite{LeungWang01,YuanChappell04,IvanovaEtAl07}. We will use the acronym ``LMP1'' (long-memory Phase~I) to refer to the family of designs assigning the estimated MTD at each cohort, regardless of whether they employ Bayesian methods. Despite their popularity among statisticians, LMP1s had struggled to enter actual practice, where the conservative `3+3' experimental protocol \cite{Carter73}, which has been repeatedly shown to possess poor MTD-selection properties \cite{Storer89,ReinerEtAl99,LinShih01}, still dominates \cite{RogatkoEtAl07}. Ivy \emph{et al.} \cite{IvyEtAl10}, on behalf of the Clinical Trial Design Task Force of the NCI's Investigational Drug Steering Committee, embrace the new designs, suggesting that \emph{``...members of the boards may not be convinced that novel designs are better for patients. In fact, they are.''}

Even as clinicians turn from skepticism to optimism, the task of constructing a comprehensive picture of LMP1 properties in theory and practice is far from complete. The available theoretical results on LMP1s are partial, and mostly involves asymptotic behavior. Azriel \emph{et al.} proved that no LMP1 design can guarantee almost-sure convergence to the MTD on the class of all dose-toxicity functions \cite{AzrielEtAl11}. Lee and Cheung \cite{LeeCheung09} provide a design tool that automatically produces a one-parameter family of curves for the continual reassessment method (CRM), upon the specification of an ``indifference interval'' around the target toxicity rate. Azriel \cite{Azriel12} proved that the conditions used by this tool indeed guarantee convergence to the specified interval -- a weaker result than converging to the MTD itself, but practically encouraging. Oron \emph{et al.} \cite{OronEtAl11ccd} show that a novel nonparametric ``interval design'' \cite{IvanovaEtAl07} is an LMP1, and prove that it converges to an interval in a very similar manner. This establishes a close asymptotic equivalence between two very different LMP1s. In Oron \emph{et al.}'s numerical examination of the convergence of one-parameter CRM and the interval design under a random sample of dose-toxicity curves, for both designs the majority of scenarios did \emph{not} meet the requisite conditions for convergence to the MTD itself. Hence, it appears that with LMP1s one must settle at best for the interval guarantee, rather than expect convergence to the MTD \cite{OronEtAl11ccd}.

Much less is definitively known regarding small-sample behavior. Despite the relatively small number of published BP1 studies, some of these have reported disturbing small-sample behavior, prompting the analysts to develop ad-hoc design modifications that might mitigate it \cite{Neunch08,RescheRigonEtAl08}. A numerical study compared the success rate of CRM designs in selecting a dose within a toxicity indifference-interval, to an ``optimal'' hypothetical experiment in which the location of each subject's toxicity-threshold with respect to the dose space is exactly known \cite{OQuigley02,PaolettiEtAl04}. One-parameter CRM performed on average very closely to the hypothetical complete-information experiment, on a class of randomly generated scenarios. Our own simulation experience (a subset of which is presented in Section~\ref{sec:numer}) suggests that several other designs can produce MTD-selection performance on par with one-parameter CRM.

%

Numerical Phase~I studies have focused almost exclusively upon ensemble-average statistics. While average values are important, in practice one does not run an \emph{ensemble} -- but rather a single experiment. A case in point is the number of cohorts treated at the MTD, a statistic we shall refer to as $n^*$. The high ensemble-average values of $n^*$ when using LMP1s have been repeatedly invoked as a decisive reason for preferring this design family \cite{RogatkoEtAl07,ZoharEtAl12}. Iasonos \emph{et al.} \cite{IasonosEtAl08}, perhaps the only study to date to present a measure of $n^*$'s variability, report large standard deviations along with these high averages (ref. \cite{IasonosEtAl08}, Table~2). Our numerical studies (Section~\ref{sec:numer}) describe the complete distribution of $n^*$ under various scenarios and designs. LMP1s suffer from alarmingly high run-to-run $n^*$ variability. The between-run variability is related to LMP1's overarching feature, namely the insistence upon treating every cohort with the estimated MTD at any given time.  The considerable operational complexity of model-based LMP1 designs, especially BP1s, often exacerbates matters.

The article is organized as follows: Section~\ref{sec:prelim} defines terminology and describes LMP1's operating principles. Section~\ref{sec:exper} presents detailed examples from published BP1 experiments. Section~\ref{sec:numer} numerically compares CRM, an ``interval design'' and a \emph{short-memory} ``Up-and-Down'' design. A general discussion ends the article.


\section{Preliminaries}\label{sec:prelim}
\subsection{Basic Terminology}\label{sec:terms}

Consider trials carried out as sequential dose-finding experiments with $n$ cohorts, indexed $c,c=1,\ldots n$, each cohort comprising of $k_c\geq 1$ subjects. Except for cohort $1$, the dose administered to cohort $c\ $ is (generally speaking) not known until all observations up to cohort $c-1$ are available. $Y_c$, the number of dose-limiting toxicities (DLT's) observed in cohort $c$, can be modeled as a Binomial random variable:

\begin{equation}\label{eq:BinomF}
Y_c \sim\textrm{Binomial}\left(k_c,F(x_c)\right),
\end{equation}
where $x_c$ is the dose administered to cohort $c$, and $F$ is the true (and unknown) underlying toxicity function, assumed to be a continuous strictly increasing cumulative distribution function (CDF) of the response-triggering dose variable $x$. In these terms, the experiment's goal is to find $F^{-1}(p)$ -- the $100p$-th percentile of $F$. This dose is known as the experiment's \textbf{ target}. In Phase~I experiments, $p$ is usually between $1/5$ and $1/3$. Doses themselves are restricted to a finite set of levels $\mathcal{D}\equiv \{d_u\},u=1,\ldots l$, with $l$ usually between $4$ and $10$. The dose level closest to $\widehat{F^{-1}(p)}$ will typically be recommended as the MTD for Phase~II. Hereafter, we denote this level as $\widehat{MTD}$.

A generic BP1 design can be described as one where in order to decide which dose to assign to the next cohort, all hitherto available observations are used to estimate $F$ via the model
\begin{equation}\label{eq:BinomG}
R_u\sim\textrm{Binomial}\left(n_u,G\left(d_u,\theta\right)\right),\ \ u=1,\ldots l,
\end{equation}
where $n_u$ is the number of available observations at dose $d_u$, $R_u$ is the number of those among the $n_u$ who exhibit toxicities, and $G$, \textbf{ the model curve}, is a CDF belonging to a parametric family $\mathcal{G}$ indexed by a parameter vector $\theta$ (which usually has a prior distribution with additional, fixed parameters). Input data to the model can be summarized as the observed sample proportions
\begin{equation}\label{eq:Fhat}
\hat{F}_u\equiv\frac{R_u}{n_u},\ \ u:\ n_u>0,
\end{equation}
which are the sufficient statistics for a nonparametric model of $F$.

According to Rogatko \emph{et al.} \cite{RogatkoEtAl07}, most BP1 experiments published through 2006 had used a one-parameter CRM model, most often of the generic form\footnote{A more sophisticated one-parameter model by Chevret \cite{Chevret93} is described in Supplement~A.}
\begin{equation}\label{eq:crm0}
G\left(d_u\right)=\phi_u^\theta,\ \ \ \phi_1<\phi_2<\ldots <\phi_l ,\ \ \phi_u\in(0,1) \ \forall u,\ \theta>0.
\end{equation}
The $\phi_u$, a sequence of constants supplied by the user, are known as the model's ``skeleton.'' This model form requires, beside the single data-estimable parameter $\theta$, the specification of $l$ fixed parameters defining the skeleton, as well as additional fixed parameters involved in $\theta$'s prior distribution.

After cohort $c$, BP1s assign the next dose to $\widehat{MTD}$, via Bayesian posterior estimation of $\hat{G}$ at the dose levels and some optimization criterion. The most common criterion is choosing the dose that minimizes $\left|\hat{G}-p\right|$, although variations exist. For example, escalation with overdose control \cite[EWOC,][]{BabbEtAl98} assigns to the dose closest to the $\alpha$ posterior quantile of $G^{-1}(p)$, with $\alpha=0.25$ often used.

Equation (\ref{eq:BinomG}) is also applicable to frequentist long-memory designs, and even to nonparametric ones. The latter directly use the $\hat{F}$, which can be viewed as a special case of $\hat{G}$. Therefore, we treat any design that allocates successive cohorts to some variation on $\widehat{MTD}$ via estimation of a model of the general form (\ref{eq:BinomG}),  as belonging to the long-memory family, regardless of whether it employs nonparametric, parametric or Bayesian methods. For this family we use the acronym LMP1, with BP1s forming a subfamily within it.

LMP1 designs were called ``designs with memory'' by some researchers \cite{OQuigleyZohar06}. Here we prefer the term \emph{long-memory} designs, in contrast with \emph{short-memory} designs that only use recent observations. However, beside their long memory LMP1s share another rather crucial trait: basing each dose allocation upon $\widehat{MTD}$. This feature prompted Fedorov \emph{et al.} to call the LMP1 family ``Best-Intention Designs'' \cite{FedorovEtAl11}. There exist Bayesian long-memory designs that employ different principles for dose assignment \cite{WhiteheadBrunier95,WhiteheadEtAl01}. Our definition of LMP1s excludes them.

\subsection{LMP1's Operating Principle}\label{sec:engine}

\begin{figure}
\begin{center}
\includegraphics[scale=0.5]{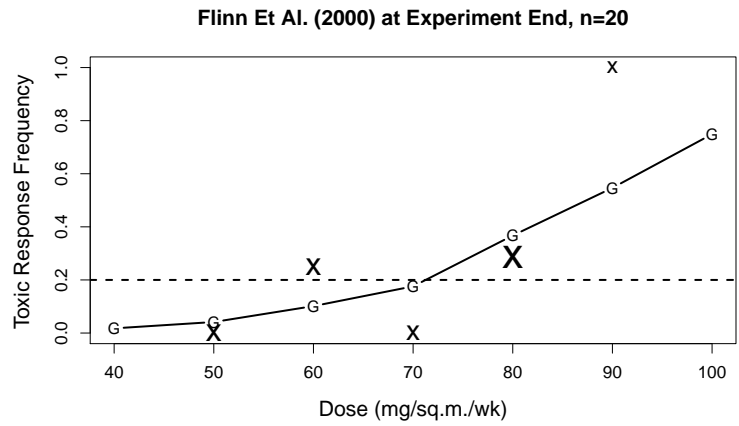}
\end{center}
\caption{The Flinn \emph{et al.} \cite{FlinnEtAl00} experiment, targeting $20\%$ toxicity (horizontal dashed line). Shown are observed toxicity frequencies (`X' marks) and the posterior model curve (connected `G' marks) at the experiment's end. `X' mark area is proportional to sample size at each dose.}\label{fig:Flinn00}
\end{figure}


The LMP1 dose-assignment process is akin to fitting a regression curve, constrained by the model family $\mathcal{G}$, through the points $\left\{\left(d_u,\hat{F}_u\right)\right\}$. These points are the `X's in Figure~1, displaying data at the end of a published CRM experiment  \cite{FlinnEtAl00}. The regression is fit by a weighted combination of the prior and the likelihood, which is itself weighted by the number of observations at each dose. The experiment's goal is finding the dose closest to the place where the true $F$ crosses the horizontal $\ y=p\ $ dashed line in Figure~\ref{fig:Flinn00}. LMP1s allocate each cohort to the best current candidate dose, according to the fitted $\hat{G}$ curve.  If more toxicities are subsequently observed at that dose, the corresponding `X' mark will move higher, pulling $\hat{G}$ with it and eventually mandating dose de-escalation, and vice versa. This is LMP1's basic self-correction mechanism, with the underlying intuition that the empirical proportions $\hat{F}$ will eventually converge to their \emph{true} $F$ values, as indeed has been recently proven for generic sequential dose-finding designs \cite{OronEtAl11ccd}.

These two elements -- self-correction in the assumed direction of target, and consistency of observed toxicity rates -- form the ``engine'' driving LMP1s. These elements are so simple, that a model $\mathcal{G}$ for $F$ is not even needed in order construct the ``engine''. For example, \emph{interval designs} \cite{IvanovaEtAl07} have no model. Instead, they mandate dose escalation if $\hat{F}$ at the current dose is below some ``tolerance interval'' around $p$, and vice versa.

If a model is used, the operating principles dictate the relationship between its slope and experimental trajectories. Shallow model curves will shift the crossing point more dramatically as $\hat{G}$ changes. Hence, they are associated with more volatile dose allocations, and vice versa for steep curves. Convex one-parameter $\mathcal{G}$ skeletons, shallow to the left and steep to the right, are rather popular in practice. They are quick to descend but more conservative when escalating. Generally, multi-parameter models can adapt the fitted slope to the observations. However, at present there is no LMP1 convergence proof for misspecified multi-parameter models.

It is important to note that the model-determined degree of volatility is unrelated to the actual rate of convergence to the MTD. The latter is paced by the convergence rate of $\hat{F}$, i.e., root-$n$ \cite{ShenOQuigley96,OronEtAl11ccd}. This is a very slow rate compared with typical Phase~I sample sizes of $10-40$ patients. If $\mathcal{G}$ correctly specifies $F$, then all data are pooled to consistently estimate $\theta$, providing the fastest possible convergence within the root-$n$ constraints. In the more likely case of misspecification, this pooling affords little help. Convergence to the MTD is then constrained by the convergence of individual $\hat{F}$'s around target. Rather often, such convergence is not guaranteed at all. As mentioned in the Introduction, the best LMP1s can guarantee without substantial restrictions on the form of $F$, is convergence to an ``indifference interval'' around $p$ which might contain several levels.


\section{Experimental Examples}\label{sec:exper}

We present in this section four published BP1 experiments. Each experiment is accompanied by a figure, in which the left-hand frame describes the experiment's trajectory -- i.e., each cohort's administered dose levels and the number of toxic and non-toxic responses observed for each, arranged in chronological order -- and the right-hand frame presents the evolution of posterior model curves. For brevity's sake, some model details are relegated to Supplement~A. Several additional experiments are decribed in Supplement~B.

\subsection{Dougherty \emph{et al.}'s Anesthesiology Experiment \cite{DoughertyEtAl00}}

\begin{figure}
\begin{center}
\includegraphics[scale=0.45]{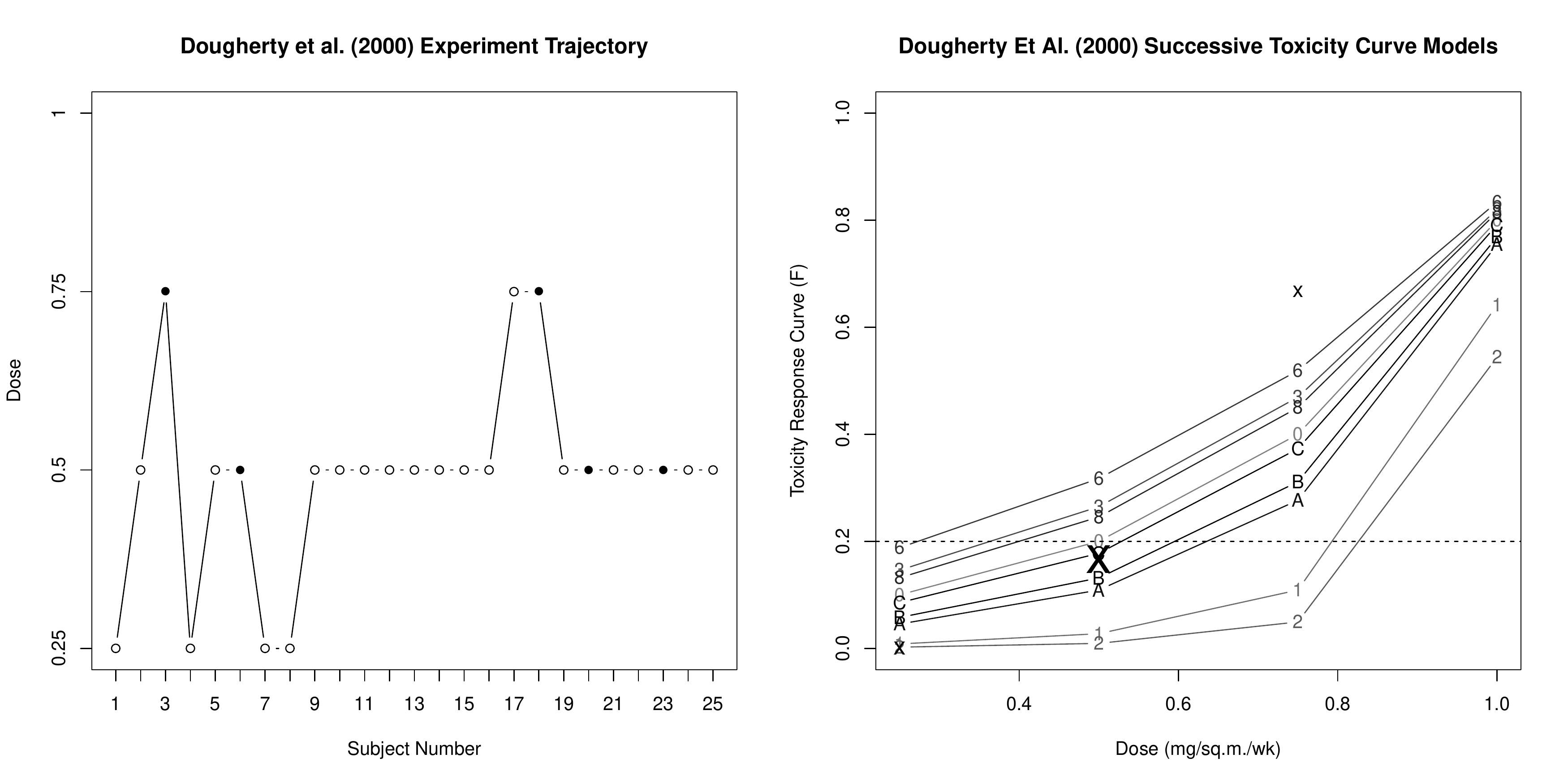}
\end{center}
\caption{Experimental trajectory and dose-response curves (right) from the Dougherty \emph{et al.} experiment \cite{DoughertyEtAl00}. In the left frame, subjects are shown in chronological order plotted against the administered levels; each empty circle represents a single negative (no-pain) response, and each filled circle represents a positive response. In the right frame, final empirical pain-rates ($\hat{F}$) are shown in `X' marks, whose size is proportional to the number of observations. The piecewise-linear curves represent posterior predictive toxicity estimates, with the number indicating the last subject before the update. The zero-symbol curve is the prior, and the symbols A, B and C stand for estimates after the 16th, 18th and 25th subject, respectively. The dashed horizontal line indicates the target response rate, in this case $0.2$.}\label{fig:doughy00}
\end{figure}

This study (Figure \ref{fig:doughy00}) was not, strictly speaking, a Phase~I trial, but rather a CRM design applied to an anesthesiology dose-finding experiment \cite{DoughertyEtAl00}. Instead of toxicity, a positive response indicates pain. The target pain rate was $0.2$, and there were $25$ patients treated one at a time. Chevret's \cite{Chevret93} one-parameter logistic model was used. There were $4$ levels in this design, with skeleton pain probabilities set at $\phi=\left(0.1,0.2,0.4,0.8\right)$. The Goodman \emph{et al.} \cite{GoodmanEtAl95} constraint, forbidding escalation by more than one level between successive cohorts, was in effect. According to its bottom line, the experiment was an astounding success: $18$ of $25$ patients were treated at the $\widehat{MTD}$ ($d_2$), with a cumulative pain rate of $3$ out of $18$ -- almost as close to target as possible ($4$ of $18$ would have been slightly closer).

\singlespacing

\doublespacing

\subsection{Pisters \emph{et al.} \cite{PistersEtAl04} and Mathew \emph{et al.} \cite{MathewEtAl04}}

\begin{figure}
\begin{center}
\includegraphics[scale=0.45]{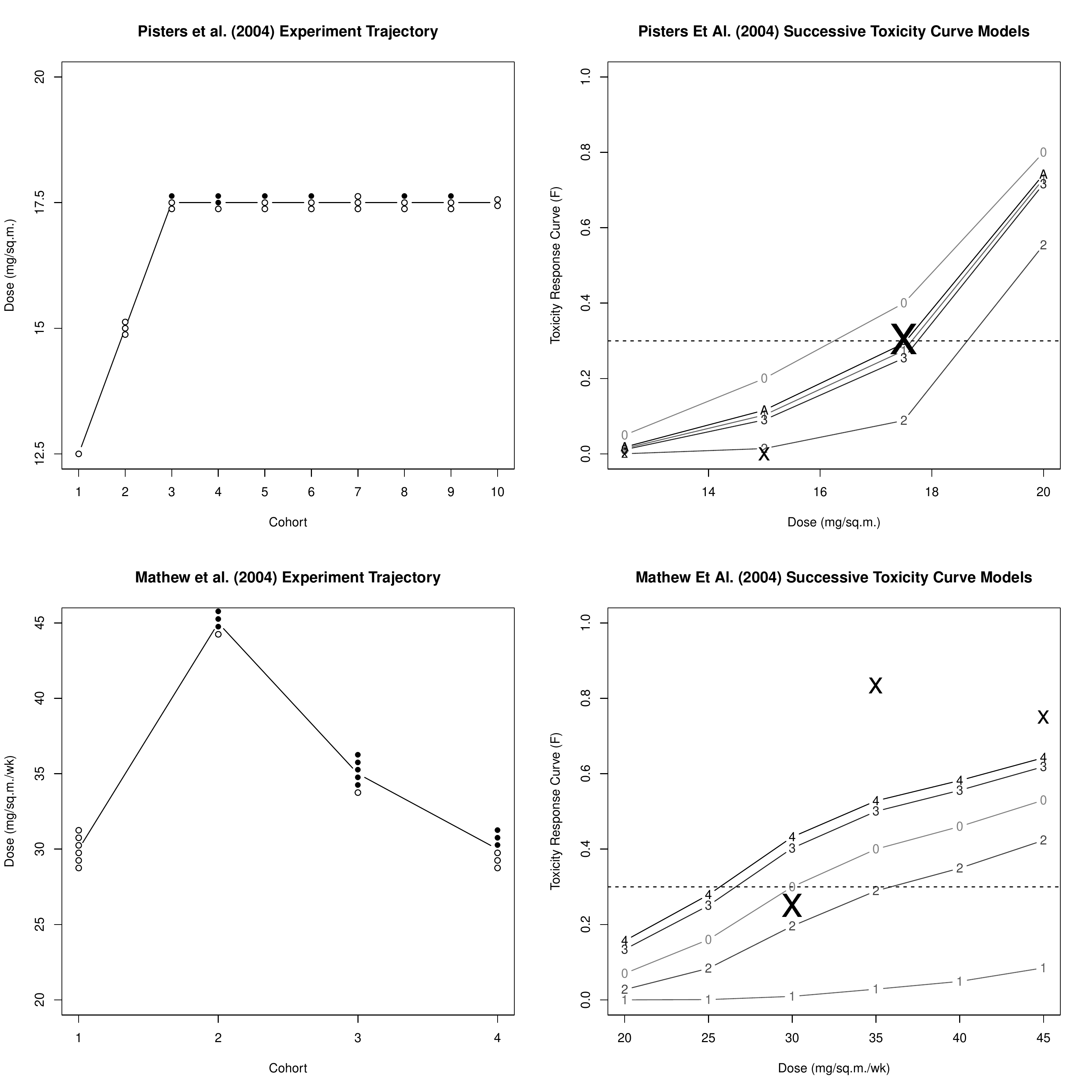}
\end{center}
\caption{The Pisters \emph{et al.} \cite[top]{PistersEtAl04} and Mathew \emph{et al.} \cite[bottom]{MathewEtAl04} experiments, using a convention similar to that of Figure \ref{fig:doughy00}. The curve with the ``A'' symbols in the top right frame indicates the final posterior after cohort $10$.}\label{fig:pistmat04}
\end{figure}

A pair of experiments conducted at the M.D. Anderson Center and published in 2004 targeted $p=0.3$, using a one-parameter `power' model CRM \cite{PistersEtAl04,MathewEtAl04}. The former followed the single-level increment constraint \cite{GoodmanEtAl95}, and had $4$ dose levels with prior toxicity probabilities nearly identical to Dougherty \emph{et al.}'s: $\phi=\left(0.05,0.20,0.40,0.80\right)$ (Figure \ref{fig:pistmat04}, top). After an unplanned single patient at $d_1$ and the first $3$-patient cohort at $d_2$ with no DLT's observed in either, {\it all}\ \   $23$ remaining patients ($8$ cohorts) were assigned $d_3$. The observed DLT rate at this dose ($7/23$) was the closest possible to target with $23$ observations; not surprisingly $d_3$ was the final $\widehat{MTD}$.

\singlespacing

\doublespacing

The story was different for the second experiment \cite{MathewEtAl04}, which neglected to follow the single-escalation constraint. The design called for six-person cohorts, and had six levels with a relatively shallow skeleton  $\phi=\left(0.07, 0.16, 0.30, 0.40, 0.46, 0.53\right)$, beginning at $d_3$ (Figure~\ref{fig:pistmat04}, bottom). After zero toxicities observed on the first cohort, allocation jumped directly to $d_6$  -- where $3$ out of $4$ toxicities forced the experimenters to cut the cohort short and de-escalate to $d_4$. At that level, $5$ toxicities out of $6$ were observed, so the experiment descended back to $d_3$, where now $3$ of $6$ experienced DLT's. This dose, with a cumulative toxicity rate of $0.25$, was recommended as the MTD; but not before half the patients in the study ($11$ of $22$) experienced DLT's. More disturbingly, a recalculation of $\hat{G}$ according to the model indicates that the final MTD estimate should have been $d_2$, with a posterior $\hat{G}=0.28$ compared to $0.43$ for $d_3$ (Figure~\ref{fig:pistmat04}, bottom right, curve marked `4'). This level had never been assigned during the experiment. Moreover, $d_2$, rather than $d_3$ should have been assigned to the last cohort as well ($\hat{G}=0.25$ and $0.40$, respectively; curve marked `3').\footnote{We inquired with the consulting statistician to this study, and he could not recall the circumstances surrounding the decisions to overrule $d_2$ with $d_3$.}

\singlespacing
\begin{figure}
\begin{center}
\includegraphics[scale=0.45]{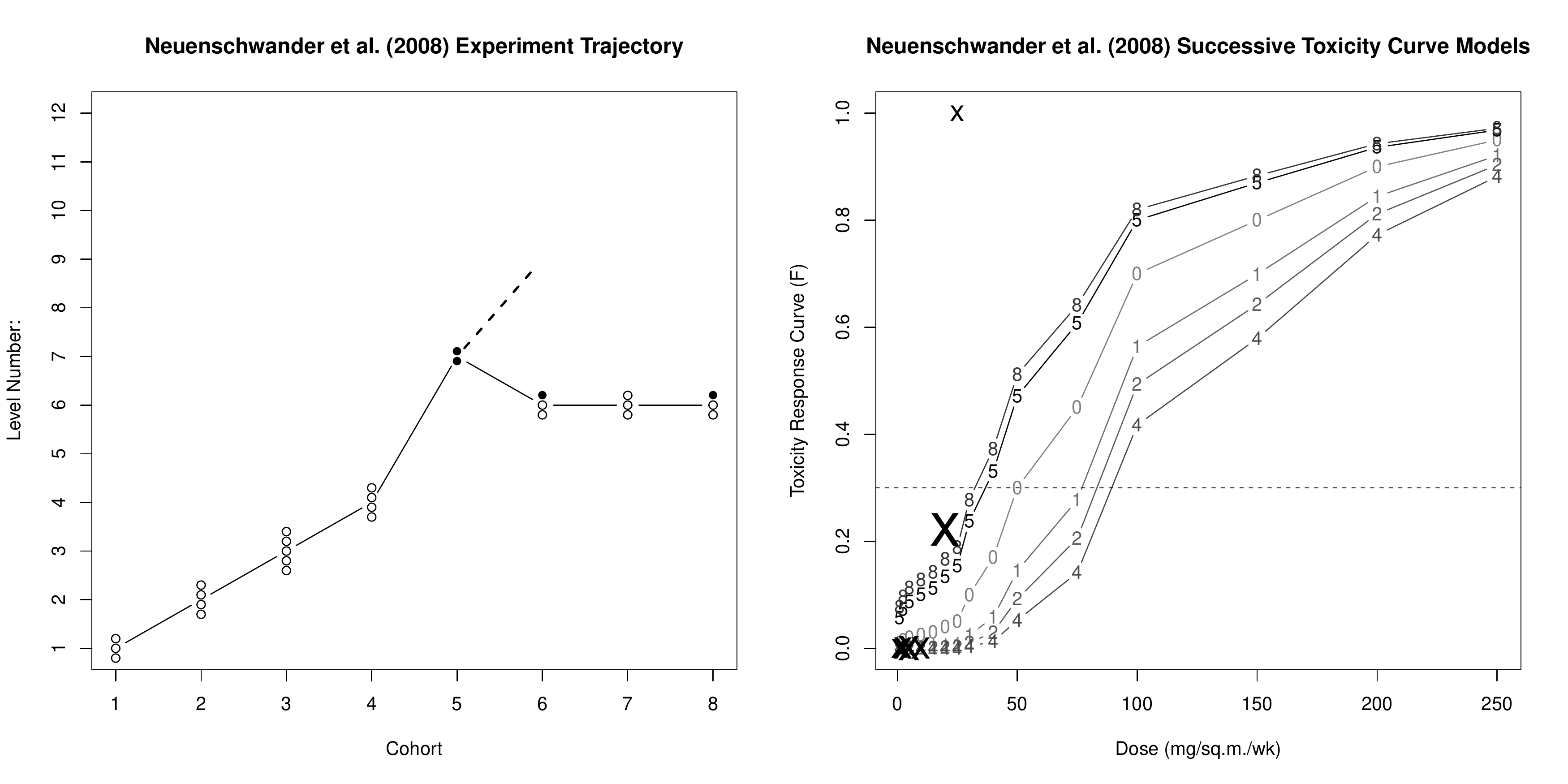}
\end{center}
\caption{Trajectory (left) and posterior model curves (right) of the Neuenschwander \emph{et al.} experiment \cite{Neunch08}. The dashed line after cohort $5$ in the left-hand frame indicates the original allocation to cohort $6$ using the one-parameter model.}\label{fig:N08}
\end{figure}
\doublespacing

\subsection{Neuenschwander \emph{et al.} \cite{Neunch08}}\label{sec:neunch}

This experiment began as a one-parameter `power' CRM, with a large number of levels, $l=15$ (Figure~\ref{fig:N08}). The starting dose was $X_1=d_{1}$, and the single-level escalation restriction was initially in effect. The predictive prior indicated $\widehat{MTD}=d_{10}$, creating an immediate tension between posterior recommendations and dose-escalation restrictions. After $4$ cohorts with $16$ patients, cumulatively, yielded no toxicities, $\widehat{MTD}$ was $d_{12}$ and researchers agreed to skip from $d_4$ to $d_7$. The next two patients both experienced DLT's, but CRM still recommended jumping from $d_7$ to $d_9$ rather than de-escalating. This type of dose-transition recommendation (i.e., escalation following toxicity or vice versa) was called ``incoherent'' by Cheung \cite{Cheung05}.\footnote{This Phase~I specific use of the term ``coherence'' can be applied to any Phase~I experiment, not just those using BP1s. It should not be confused with Bayesian coherence \cite{Lindley83}.} Cheung proved that if the current dose is also the $\widehat{MTD}$, one-parameter CRM designs cannot make ``incoherent'' recommendations. However, if $X_1\neq\widehat{MTD}$, then ``incoherence'' might be encountered during the experiment's initial phase until the gap between the two is closed -- which is what happened to the experiment in question.

After the ``incoherent'' recommendation, the trial was put on hold, and intensive simulation and theory work was carried out in order to modify the design \cite{Neunch08}. The authors replaced the one-parameter model with a two-parameter logistic, and modified the decision rule to penalize toxicity more heavily. These changes resulted in $d_6$ (i.e., a one-level de-escalation) being recommended for the trial's continuation. All $3$ remaining cohorts were administered that dose, which eventually became the final $\widehat{MTD}$ with $2$ toxicities observed on $9$ patients. Interestingly, even at the experiment's end the original one-parameter model still estimated $\widehat{MTD}=d_8$, rather than $d_6$ (Figure~\ref{fig:N08}, right, curve marked `8').




\section{Numerical Demonstrations}\label{sec:numer}

\subsection{Overview and Methods}

In this section, we numerically examine some aspects of LMP1 behavior, compared with short-memory designs belonging to the {\it ``Up-and-Down''}\ \ family \cite{DixonMood48}. The simulated sample size was $n=32$ subjects in cohorts of $2$ on $l=6$ dose levels, for all designs. Several $6$-tuples of $F$ values on $\mathcal{D}$, hereafter called \textbf{scenarios}, were chosen using different parametric curve forms for $F$; hence the scenario names as they appear later. For each scenario, $M=1000$ sets of (pseudo-)random toxicity thresholds, representing patients with different toxicity sensitivities, were drawn from that scenario's exact parametric model for $F$. For each design, each of the $M$ simulated experiments (hereafter: \textbf{runs}) compared each cohort's assigned doses to its toxicity thresholds, and subsequently applied the design's transition rules. The group of $M$ runs drawn from the same distribution is known as an \textbf{ensemble}.  Each design was tested under the exact same conditions, encountering the same toxicity thresholds in the same order as the other designs. The experiment's formal target was $p=0.3$ throughout the simulation. We show results from six scenarios, calibrated so that each of the $6$ dose levels was the true MTD for one scenario.

Up-and-Down (U\&D)  is often conflated by authors in the field \cite[e.g., by][]{RogatkoEtAl07,Neunch08} with the 3+3 protocol. However, the two diverge in several important respects -- first and foremost, the fact that U\&D is a statistical design with clear theoretical properties, while 3+3 is not. See Supplement~C for a more detailed list of differences. U\&D designs generate Markov chains over the dose space, with visit frequencies peaking near $F^{-1}(p)$ \cite{Derman57,Tsutakawa67}. Their convergence rate to this asymptotic behavior is geometric. Recent methodological work on U\&D has explored their properties as nonparametric designs, and developed novel variations, extensions and estimation methods \cite{DurhamFlournoy95,Gezmu96,StylianouFlournoy02,IvanovaEtAl03,OronHoff09}. However, the overall number of recent methodological U\&D publications is at least an order of magnitude smaller than analogous work on LMP1s.

For the simulations illustrated here, we used a group U\&D design \cite{Tsutakawa67,GezmuFlournoy06}. The allocation rule is
$$
\left.\begin{array}{l}
\textrm{1.  Start at a pre-designated dose, and treat cohorts of size 2;}\\
\textrm{2. After cohort $c$ is treated at } X_c=d_u, \textrm{ set } X_{c+1} \textrm{ to}
\left\{\begin{array}{ll}
d_{u+1} & \textrm{If 0 of 2 are toxicities,} \\
d_{u-1} & \mathrm{otherwise.} \
\end{array}\right.
\end{array}\right.
$$
Boundary conditions, as usual, replace non-existent levels ($d_0$ or $d_{l+1}$) with the existing levels $d_1,d_l$, respectively, whenever the former are mandated. The design converges to an asymptotic allocation distribution peaked near $F^{-1}(0.29)$.

For CRM, the simulation used the ``power'' model with a skeleton similar to that of Flinn \emph{et al.} (Figure~\ref{fig:Flinn00}). Those authors' skeleton was $\phi=\left(0.05,0.10,0.20,0.30,0.50,0.65,0.80\right)$ with $l=7$, and ours is $\phi=\left(0.05,0.11,0.22,0.40,0.60,0.78\right)$ with $l=6$. The prior on $\theta$ was log-Normal, the one most commonly used in practice, and was calibrated so that initial responses be ``coherent'' in the sense defined in Section~\ref{sec:neunch}. The single-level escalation constraint was universally used, in both the upward and downward directions.

Another LMP1 examined here is Ivanova \emph{et al.}'s  nonparametric ``cumulative cohort design'' (CCD) \cite{IvanovaEtAl07}. CCD is an interval design, which -- as mentioned in the Introduction -- has a convergence behavior rather similar to CRM. Apart from that, CCD is perhaps the LMP1 design type most different from CRM: it repeats the same dose $d_u$ as long as $\hat{F_u}$ falls inside a tolerance interval around $p$, escalates if $\hat{F_u}$ is below the interval and vice versa . Here we used the interval $(0.2,0.4)$, recommended in \cite{IvanovaEtAl07} for $l=6$.

All runs started at $d_2$. The code and subsequent analysis were implemented in R \cite{RLang}. Further details (simulation scenario curves, etc.) appear in Supplement~D.

\subsection{Between-Run Variability and the Order Effect}\label{sec:ordersim}

Dose-finding simulation summaries are usually statistics of \emph{average} ensemble performance, for example the proportion of runs for which the true MTD was found by various designs, or $\overline{n^*}$ -- the ensemble average number of cohorts administered the true MTD. Many LMP1 designs tend to perform well on the latter statistic which also happens to be one of U\&D's weakest aspects, being a random-walk design that inevitably spreads allocations over several levels around $F^{-1}(p)$   .

\begin{figure}
\begin{center}
\includegraphics[scale=0.7]{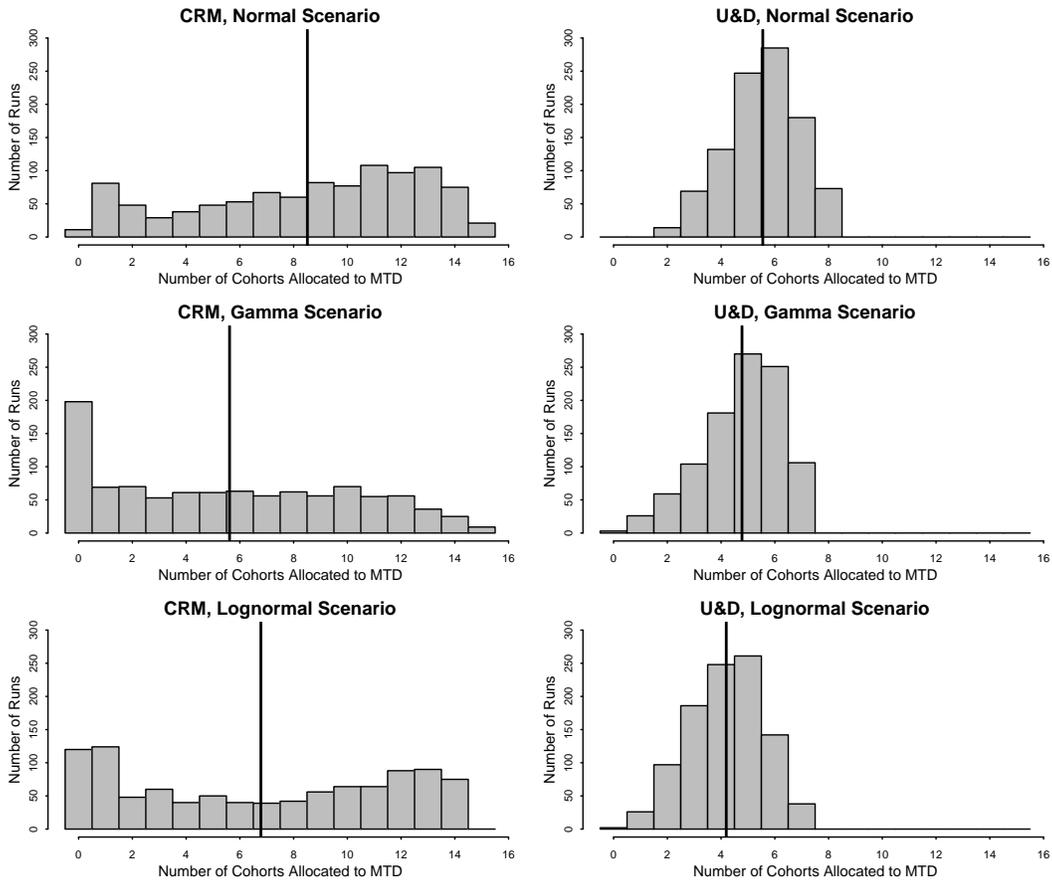}
\end{center}
\caption{Between-run and between-scenario variability. The histograms depict the ensemble distribution of $n^*$, excluding the first cohort. The ensemble size is $1000$ runs. Scenarios are Normal (top), Gamma (middle) and Lognormal (bottom); designs are CRM one-parameter `power' (left) and GU\&D (right), both with cohort size $2$. The runs were $16$ cohorts long, starting at $d_2$.}\label{fig:BestFig}
\end{figure}

Rather than just report the average, Figure~\ref{fig:BestFig} displays the ensemble \emph{distribution} of $n^*$ (excluding the arbitrary first cohort), enabling a glimpse into run-to-run variability. The ensemble average $\overline{n^*}$ is visible as the bold vertical line in the middle of each histogram. One-parameter CRM (left) is compared here with group U\&D \cite{Tsutakawa67,GezmuFlournoy06} (right). Shown are three of the scenarios (top to bottom).

The most dramatic feature in Figure~\ref{fig:BestFig} is CRM's between-run variability. Even under the Normal scenario (top left), where the ensemble mode is at a spectacular $11$ MTD-allocated cohorts out of $15$ and the average is around $8.5$ cohorts, $14\%$ of the CRM runs ended with $n^*\leq 2$ -- meaning that fewer patients were treated at the true MTD than would have under a fixed uniform allocation rule across the dose levels. In the Gamma scenario (middle left), CRM's most common $n^*$  outcome allocates \emph{zero} cohorts to the MTD during the experiment. It should be noted that for the Gamma scenario, the MTD was actually the starting dose ($d_2$), meaning that in one-fifth of the runs CRM immediately veered away from its starting dose, never to return -- despite $d_2$ being the correct MTD. Finally, the log-Normal scenario (bottom left) generates strongly divergent behavior, with very low or very high values of $n^*$ more common than intermediate outcomes.

With U\&D (Figure~\ref{fig:BestFig}, right-hand frames), between-run and between-scenario differences are far smaller. Due to its random-walk nature, group U\&D cannot allocate more than roughly half the cohorts to any single level except on the boundary. However, in all scenarios the modal outcome is reasonably close to this limit at $5-6$ cohorts per run, with the vast majority of runs producing $n^*$ values within $\pm 2$ of the mode.

CRM's $n^*$ variability indicates a sensitivity to variations in the input data. Sensitivity to model fit definitely exists, as can be deduced from the differences between scenarios. However, this does not explain the considerable within-scenario sensitivity between different runs, since these runs were all drawn from the same $F$. Two remaining sources of variability are:
\begin{enumerate}
\item Each run's empirical sample moments. For example, if a single run's set of 32 thresholds is uncharacteristically high (low) on the average, the run will likely have a smaller (larger) number of toxic responses, respectively, unless the design self-corrects upward (downward). This is analogous to two groups of patients who by random chance are on the average more or less liable to respond toxically to the treatment, despite belonging to the same patient population.
\item The \emph{order} in which toxicity-thresholds are encountered during the run. For example, would a group of low thresholds followed by a group of high ones, generate different behavior than if the two groups' order was reversed?
\end{enumerate}

\begin{figure}
\begin{center}
\includegraphics[scale=0.7]{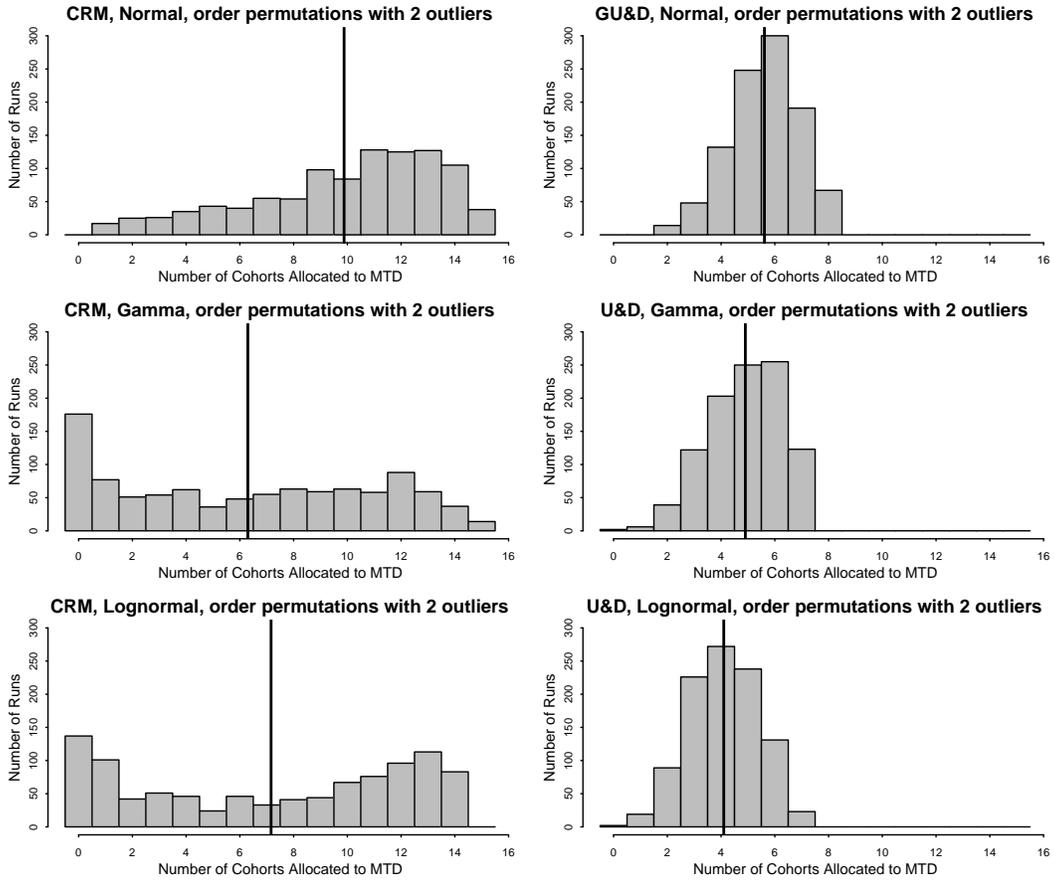}
\end{center}
\caption{Similar to Figure \ref{fig:BestFig}, except that rather than draws out of a simulated distribution, the runs are order permutations of the same set of $32$ thresholds, as described in the text.}\label{fig:ScrFig}
\end{figure}

To help pinpoint which of the two sources is more responsible for CRM's $n^*$ variability, we replaced the randomly-generated thresholds with fixed sets. For each scenario, a ``perfect set'' consisting of the percentiles $F^{-1}(1/33),\ldots F^{-1}(32/33)$, was slightly modified by ``knocking out'' two thresholds in the vicinity of $F^{-1}(p)$, one on each side, and replacing them with replicas of $F^{-1}(1/33)$ and $F^{-1}(32/33)$, respectively (the original ``perfect set'' would be unrealistically well-behaved). We then generated $M=1000$ runs from each scenario, each run using the exact same set of thresholds, but with the \emph{order} in which they appear randomly permuted. Figure~\ref{fig:ScrFig} shows the distributions of $n^*$ from these runs; it is impressively similar to Figure~\ref{fig:BestFig}. This establishes that CRM's run-to-run variability in $n^*$ is driven primarily by variations in sampling order, i.e., the order in which participants enter the experiment.

\begin{figure}
\begin{center}
\includegraphics[scale=0.7]{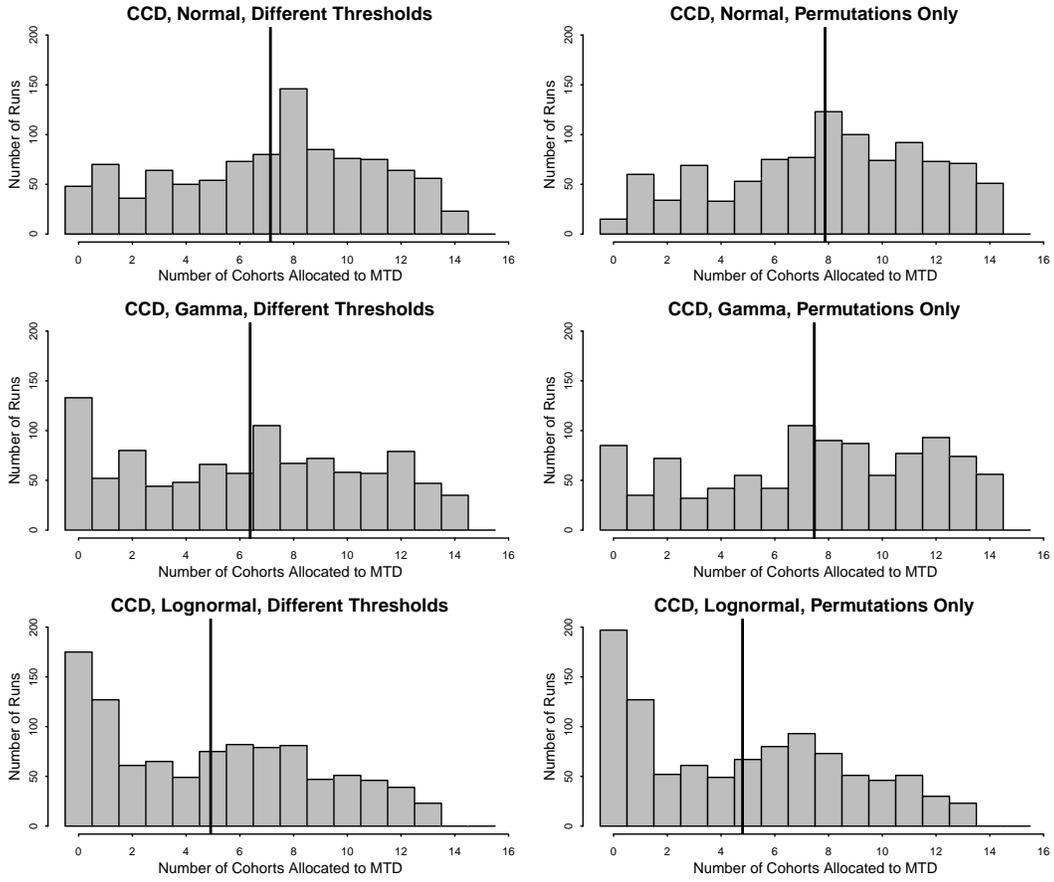}
\end{center}
\caption{Distribution of $n^*$ using the exact random draws of Figs. \ref{fig:BestFig} (left) and \ref{fig:ScrFig} (right), under the nonparametric interval design CCD \cite{IvanovaEtAl07}.}\label{fig:CCDFig}
\end{figure}

Variability in $n^*$ and sensitivity to sampling order are properties of all LMP1 designs, not just Bayesian ones. Figure~\ref{fig:CCDFig} repeats the same exercise of Figures~\ref{fig:BestFig}--\ref{fig:ScrFig}  (pseudorandom draws, then permutations of a fixed threshold set), using the nonparametric CCD for dose allocations. Between-scenario variability in $n^*$ is smaller than with CRM, but between-run variability in each scenario is as great or greater.

\subsection{MTD-Selection Performance and Effect of Prior}

Table \ref{tbl:bulk} presents the percent of runs in which the MTD was correctly selected for the three methods under each scenario, halfway through the experiment (left three columns) and at its end (right three columns). We chose scenarios where the MTD is unambiguous: its true $F$ value is always very close to $0.3$, and the $F$ values of neighboring levels are no closer than approximately $0.2$ or $0.4$ (see details in Supplement~D). For U\&D and CCD estimation we used a variant of isotonic regression, recently recommended as a robust and relatively efficient choice \cite{StylianouFlournoy02}. If the $\hat{F}$ are monotone increasing, then this estimator simply chooses the level with $\hat{F}$ value closest to $0.3$ as the MTD. If there is a monotonicity violation, the violating values are replaced by a single weighted average, and the level closest to the MTD is determined by linear interpolation between the new values \cite[Section~3.3]{Oron07}. The code used for isotonic-regression estimation is available in Supplement~F.

\begin{table}
\caption{Bulk performance comparison between ``power'' CRM, CCD and group U\&D. For each of six scenarios, compared are the proportion of runs in which the correct MTD was selected, after $8$ (left) and $16$ (right) cohorts, respectively. CRM is estimated as the next dose allocation; U\&D and CCD were estimated using centered isotonic regression \cite[Section~3.3]{Oron07}. All numbers are in percents.}\label{tbl:bulk}

\begin{tabular}{lcccccccc}
\\
\toprule
& \multirow{2}{1.2cm}{MTD Level} & \multicolumn{3}{c}\textbf{ After 8 Cohorts} & & \multicolumn{3}{c}\textbf{ After 16 Cohorts} \\
Scenario & & CRM & CCD & U\&D & &  CRM & CCD & U\&D \\
\midrule
``Uniform'' & 1 & 50.2 & 57.1 & 54.0 & & 62.1 & 64.1 & 60.8 \\
``Gamma'' & 2 & 36.6 & 44.2 & 40.8 & & 47.4 & 53.2 & 51.2 \\
``Normal'' & 3 & 57.8 & 54.2 & 56.4 & & 67.5 & 67.1 & 63.0 \\
``Lognormal'' & 4 & \textbf{46.7} & 34.0 & 33.0 & & \textbf{59.3} & 46.2 & 48.4 \\
``Weibull'' & 5 & 39.0 & \emph{28.1} & 38.1 & & 47.2 & 42.6 & 45.0 \\
``Logistic'' & 6 & 26.0 & 30.0 & 32.2 & & \emph{29.3} & 48.5 & \textbf{ 54.6} \\
\bottomrule
\end{tabular}
\end{table}

In the table, we used a visual convention of emphasizing only outcomes for which the design in question is substantially different ($5\%$ or more) from the others. Stronger results are in boldface and weaker results in italics. Overall, the performance differences between these three very different designs are remarkably small. It is actually CRM that falls most conspicuously behind under the `Logistic' scenario (bottom row), in which it shows nearly no improvement during the experiment's second half.

Table~\ref{tbl:prior} summarizes the MTD-selection performance of the same CRM skeleton with 3 different sets of prior parameters. The prior used to produce Figures~\ref{fig:BestFig}-\ref{fig:ScrFig} and Table~\ref{tbl:bulk} is labeled ``A''. It represents a modest amount of scientific knowledge and priorities: it assumes the middle of the dose range is somewhat more likely to contain the MTD, and the highest doses are less likely or desirable than the lowest ones (Table~\ref{tbl:prior}, left column). Prior~B, which encourages dose escalation (e.g., $d_5$ has more prior-predictive weight than $d_2$ or $d_3$), is commonly recommended by CRM researchers as ``uninformative''. It is the default prior in the \texttt{`crm'} function in the R package \texttt{`dfcrm'} described in Cheung's recent book \cite{Cheung11}. Prior~C reflects a strong belief that the MTD is in the lower half of the dose range, or (equivalently) a reluctance to prefer higher doses until overwhelming evidence has accumulated. All priors used the log-Normal distribution.

\begin{table}
\caption{Similar to Table \ref{tbl:bulk}, but only with CRM, using the same skeleton and three different priors labeled A, B and C. The first three columns show each prior's predictive $\widehat{MTD}$ distribution.}\label{tbl:prior}

\begin{tabular}{lcccccccccc}
\\
\toprule
& \multicolumn{3}{c}{\textbf{ Prior Weight}} &  \multicolumn{3}{c}\textbf{ After 8 Cohorts} & & \multicolumn{3}{c}\textbf{ After 16 Cohorts} \\
Scenario/MTD & A & B & C  & A & B & C &  & A & B & C \\
\midrule
``Uniform''/$d_1$ & 0.25 & 0.26 & 0.33  &50.2 & 53.3 & 50.2 & & 62.1 & 63.1 & 64.5 \\
``Gamma''/$d_2$& 0.14 & 0.10 & 0.22 &  36.6 & 34.4 & \textbf{ 44.3} & & 47.4 & 45.9 & \textbf{ 54.0} \\
``Normal''/$d_3$& 0.20 & 0.15 & 0.25  & 57.8 & \emph{52.6} & \textbf{ 63.2} & & 67.5 & 66.0 & \textbf{ 72.6} \\
``Lognormal''/$d_4$ & 0.22 & 0.18 & 0.16  & 46.7 & 46.6 & 45.0 & & 59.3 & 55.4 & 56.4 \\
``Weibull''/$d_5$& 0.14 & 0.17 & 0.04 &  39.0 & 39.9 & \emph{23.0} & & 47.2 & 50.6 & \emph{29.7} \\
``Logistic''/$d_6$& 0.05 & 0.15 & 0.002  & 26.0 & \textbf{ 34.6} & \emph{0.0} & & 29.3 & \textbf{ 41.0} & \emph{7.2} \\
\bottomrule
\end{tabular}
\end{table}

Overall, the performance variability when using the same CRM model with different priors is as great or greater than the variability  \emph{between} methods seen in Table~\ref{tbl:bulk}. The relative performance in Table~\ref{tbl:prior} mirrors the MTD's relative prior-predictive weight under each prior, or more precisely: each level's predictive weight compared with its immediate neighbors. The performance improvement from 16 to 32 subjects is around $10\%-15\%$ in most scenarios regardless of prior; however, it is substantially slower under the Weibull and Logistic scenarios -- the two scenarios for which this model fails to converge to the MTD.

\subsection{``Settling'' and Estimation Success}\label{sec:settl}

The phenomenon of LMP1 experiments settling fairly early on a single dose is well-known; see, e.g., the first two experiments in Section~\ref{sec:exper}. O'Quigley \cite{OQuigley06}, Rogatko \emph{et al.} \cite{RogatkoEtAl07}, and many others see it as a strength. The rationale is that such a settling indicates the LMP1 self-correction mechanism needs little further information to determine the MTD. No LMP1 convergence proof uses any mechanism faster than root-$n$, and therefore it is hard to argue that model estimates based on a sample of 10-30 binary observations, are already guaranteed to be close to their true asymptotic values. However, the question remains whether early settling in an LMP1 experiment is an encouraging sign for good outcomes, or not.

\begin{figure}
\begin{center}
\includegraphics[scale=0.7]{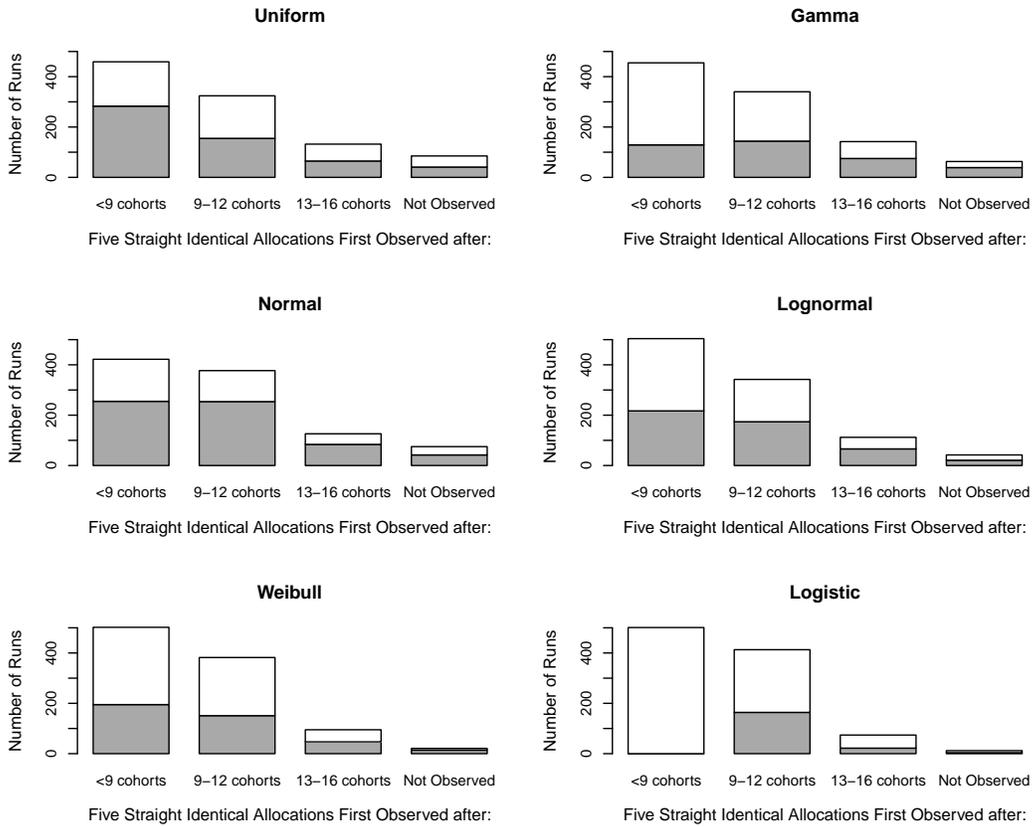}
\end{center}
\caption{CRM estimation performance, by scenario and ``settling''. Bar length is proportional to the number of runs in each ``settling'' stage. The shaded portion represents those runs ``settling'' on the true MTD at that time.}\label{fig:StagEst}
\end{figure}

Here we consider a run to have ``settled'' once the same dose has been assigned 5~\emph{consecutive} times (excluding the arbitrary starting dose). Some LMP1 studies had suggested a similar settling criterion as a stopping rule \cite{ZoharChevret01}. More sophisticated stopping-rule approaches such as calculating the posterior probability of future dose transitions, are not far removed from this simple rule. Figure~\ref{fig:StagEst} divides the bulk summaries of CRM MTD-selection performance (with Prior~A) in each scenario into four groups, according to the time at which settling is first encountered - before the ninth cohort; after 9-12~cohorts; after 13-16~cohorts; or not at all. Recall that the simulation had 16~cohorts of size~2. Bar lengths are proportional to group sizes, and the shaded regions represent the runs pointing to the correct MTD at the time of settling.

While (as shown in Figure~\ref{fig:BestFig} and Table~\ref{tbl:bulk}) CRM performance varies widely between scenarios, Figure~\ref{fig:StagEst} presents little evidence that early settling is associated with better $\widehat{MTD}$ performance; if anything, the contrary. Note however that CRM's settling behavior itself is remarkably uniform across scenarios: under all scenarios, roughly half the runs encounter five consecutive identical allocations by cohort~8, and $80\%-90\%$ of runs display this phenomenon by cohort~12. All in all, Figure~\ref{fig:StagEst} suggests that early settling is a universal CRM design side-effect, rather than a sign of quick convergence to the true MTD.

\subsection{Simulations with Random $F$}\label{sec:rand}

The simulations described above follow the common approach used in the field: fix a handful of scenarios for $F$, then simulate ensembles of random samples from each scenario. This approach is useful for isolating sources of variability, but is deficient as a performance evaluation tool. Inevitably, any small set of consciously-selected scenarios is favorable to some methods and unfavorable to others. Moreover, such simulations under-represent the amount of variability encountered in practice.

For this purpose, and also to establish the broadness of the numerical results of Figures~\ref{fig:BestFig}-\ref{fig:StagEst}, we present a brief summary of simulation results under an ensemble of random scenarios of $F$, using the methodology presented in ref. \cite{OronEtAl11ccd}. Random-$F$ simulations better approximate real applications, in which each experiment examines the effect of a different treatment for a different disease, on different patient populations.

Rather than the relatively generous $n=32$ of the fixed-scenario simulations, the random-scenario simulations used $n=25$, more in line with sample sizes of experiments described in this article. Simulated ``patients'' were treated one by one. We used $l=7$ and $l=4$ levels, enabling us to directly incorporate the model skeletons of Flinn \emph{et al.} \cite{FlinnEtAl00} and Pisters \emph{et al.} \cite{PistersEtAl04}, respectively. The former was used with the default Cheung prior, while the latter used the published prior that had mean 0 and variance $1.8$. Cheung's \texttt{`crm'} function was used to calculate dose transitions and final estimates. The target toxicity rate $p$ was left unchanged at $0.3$. For the CCD interval design, we used a width of $\pm 0.1$ as before for both values of $l$.

The leading U\&D design for one-at-a-time treatments and $p\approx 0.3$ is ``$k$-in-a-row'' \cite{IvanovaEtAl03,OronHoff09}, used extensively in sensory studies under the name ``forced-choice fixed staircase''. Its allocation rule is
$$
\left.\begin{array}{l}
\textrm{1. Start at a pre-designated dose, and treat one patient at a time;}\\
\textrm{2. After patient $c$ is treated at } X_c=d_u, \textrm{ set } X_{c+1} \textrm{ to}
\left\{\begin{array}{ll}
d_{u+1} & \textrm{If } Y_c=Y_{c-1}=0 \textrm{ and } X_c=X_{c-1}=d_u \\
d_{u-1} & \textrm{If } Y_c=1 \\
d_u & \textrm{Otherwise.}
\end{array}\right.
\end{array}\right.
$$
In words, escalation is mandated only after two consecutive non-toxicities at the current dose, while de-escalation is mandated after every toxicity. Besides the usual boundary conditions, there is a start-up condition preventing escalation before 2 observations are available. This design is closely related to the cohort-$2$ design used above \cite{OronHoff09}, and similarly converges to an allocation distribution peaked near $F^{-1}(0.29)$.

Ensemble size was $2000$ runs for each value of $l$, each run having a different, randomly-generated $F$. The ensembles are stratified samples from a larger number of random scenarios, to ensure a uniform distribution of true-MTD levels (except for somewhat lower counts on boundary levels). The $l=7$ runs began at $d_2$, and $l=4$ runs at $d_1$. Further simulation details are in Supplement~D, and the R function for generating random scenarios is in Supplement~F.

\begin{figure}
\begin{center}
\includegraphics[scale=0.65]{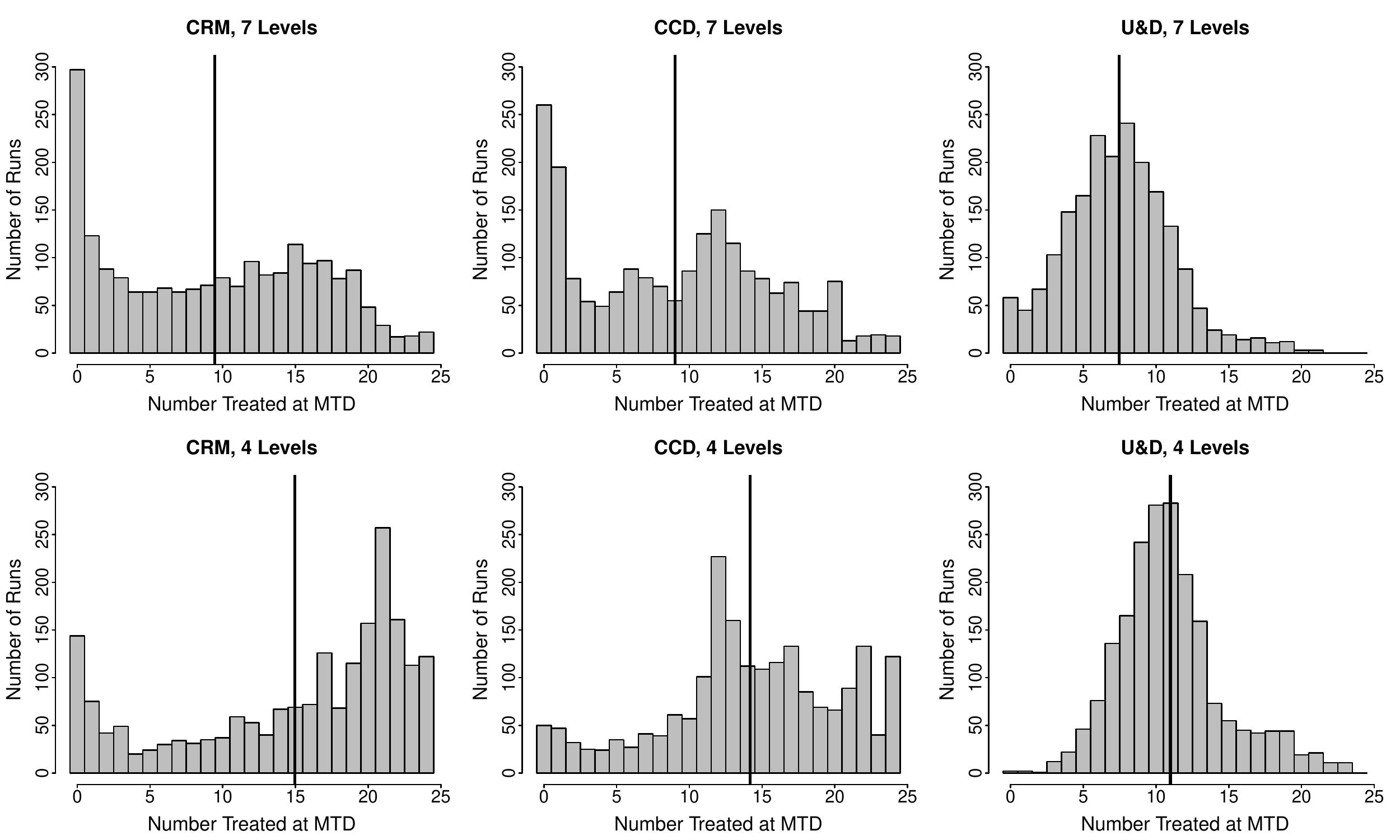}
\end{center}
\caption{Similar to Figures~\ref{fig:BestFig}--\ref{fig:CCDFig}, using data from the random scenario simulation described in the text.}\label{fig:RandFig}
\end{figure}

Figure~\ref{fig:RandFig} displays the $n^*$ distributions for CRM (left), CCD (center) and U\&D (right). With $l=7$ (top), the patterns are similar to Figures~\ref{fig:BestFig}--\ref{fig:CCDFig}, except that between-run variability tends to increase for all designs. With $l=4$ (bottom) there is a marked improvement, more dramatically for the LMP1s. CCD's peak moved to around $n/2$, suggesting a dominance of oscillating transitions between two adjacent levels, the MTD being one of them.  Some of the improvement from $l=7$ to $l=4$ is due to a doubling in the number of scenarios whose MTD is a boundary level, especially $d_1$ which is the easiest to identify. But a large part is simply due to the improved MTD-detection ability with fewer levels.

\begin{table}
\caption{Performance comparison between CRM, CCD and ``$k$-in-a-row'' U\&D, for the random-scenario simulation with $l=7$ levels (left) and $4$ levels (right), $n=25$ and $M=2000$ runs. All numbers are in percents.}\label{tbl:ranbulk}

\begin{tabular}{p{4.5cm}rrrrrrr}
\\
\toprule
&  \multicolumn{3}{c}{$\mathbf{\textit{l}=7}$} & & \multicolumn{3}{c}{$\mathbf{\textit{l}=4}$} \\
Outcome &  CRM & CCD & U\&D & &  CRM & CCD & U\&D \\
\midrule
MTD-Selection Success  & $53.0$ & $51.4$ & $51.3$ & & $75.2$ & $78.0$ & $76.5$ \\
\midrule

High-$n^*$ Runs & $43.3$ & $39.9$ & $\textit{11.9}$ & & $71.0$ & $73.1$ & $\textit{36.3}$ \\
Low-$n^*$ Runs&  $29.4$ & $29.4$ & $\mathbf{13.7}$ & & $\textit{17.7}$ & $10.7$ & $\mathbf{4.3}$ \\
\midrule
High-Toxicity Runs  & $5.1$ & $\textit{10.3}$ & $\mathbf{2.7}$ & & $4.8$ & $7.0$ & $5.8$ \\
Runs with ``Incoherence'' & $0.0$ & $\textit{86.6}$ & $0.0$  & & $0.0$ & $\textit{73.5}$ & $0.0$  \\
\bottomrule
\end{tabular}
\end{table}

Table~\ref{tbl:ranbulk} presents summary statistics, bolding over-performing outcomes and italicizing under-performing ones. Overall MTD-selection performance (top row) is remarkably similar across designs, and improves quite substantially from $l=7$ (left) to $l=4$ (right). Note that U\&D succeeds in matching LMP1 preformance, despite being handicapped by waiting 2 patients for each dose escalation (the LMP1s are allowed to escalate at each turn). Interestingly, CRM falls somewhat behind with $l=4$. The cause was traced to the model skeleton: the skeleton's jump from $0.4$ at $d_3$ to $0.8$ at $d_4$ is disproportionately large, and the predictive prior $\widehat{MTD}$ distribution is an uneven $(0.33,0.17,0.28,0.22)$. Changing the skeleton's $d_4$ value to $0.6$ and tightening the prior variance to produce a nearly-uniform predictive distribution, increased the MTD-selection rate by $2\%$ and reduced the proportion of low-$n^*$ runs by $3\%$.

Speaking of $n^*$, we replaced its ensemble average with two threshold-based summaries. The desirable-outcome statistic (second row) is the proportion of runs in which at least half of allocations $2-25$ were assigned to the true MTD. The detrimental outcome of low-$n^*$ runs (third row) was defined as the proportion of runs in which the number allocated to patients $2-25$ is smaller than $(n-1)/l$, the number obtained by simple non-sequential uniform dose allocation. The two LMP1s perform very similarly on these statistics, and the contrast between them and U\&D is stark: for LMP1s and $l=7$, only one-quarter of runs have an intermediate value of $n^*$, with the high and low tails divided roughly 4 to 3, respectively. For U\&D and $l=7$, three-quarter of runs have an intermediate $n^*$, with the tails divided roughly equally. The high/low ratios substantially improve for all designs with $l=4$.

In addition, we examined the proportion of runs with many toxicities observed on patients $2-25$ ($>9$ toxicities with $l=7$, $>10$ with $l=4$; fourth row). With $l=7$, U\&D is safest while CCD somewhat underperforms having $>10\%$ high-toxicity runs. With $l=4$, the performance is more similar across designs.

Runs containing an ``incoherent'' transition (escalation following toxicity or vice versa) were also counted (bottom row). While U\&D is hard-wired to disallow such transitions, and the two CRM skeletons appear to be well-calibrated to prevent them during the first few cohorts until the experiment reaches $\widehat{MTD}$ (see Section~\ref{sec:neunch}), CCD displays at least one ``incoherent'' transition in most runs, doubtlessly related to its tendency to sometimes oscillate between adjacent levels.


\section{Discussion and Recommendations}\label{sec:theory}

Our simulations indicate that the leading Bayesian Phase~I method -- one-parameter CRM -- delivers MTD-selection success rates similar to those of a simpler nonparametric LMP1 and the even simpler ``Up-and-Down'' designs. Finding the MTD is usually Phase~I's main goal. Given the general preference for parsimony in science and in statistical modeling, this raises the question whether LMP1s in general and BP1s in particular should still be recommended by statisticians.

LMP1s do maintain an advantage in the expected number of patients treated at the MTD, approximated by the simulation ensemble means $\overline{n^*}$. However, this comes at the price of dramatically increasing $n^*$'s variability, in particular the probability of having even fewer subjects treated at the MTD than by random chance or fixed allocation. It is unclear whether clinicians, once cognizant of this tradeoff, will see the increase in $\overline{n^*}$ as worth the added risk and complexity, especially since the MTD-selection success rate does not substantially improve. We suggest that studies of future designs always include an examination of $n^*$'s distribution, and report the ensemble proportions of various adverse outcomes as done in Table~\ref{tbl:ranbulk}. The habit of reporting only the ensemble average for non-binary outcomes such as $n^*$ masks the variability between simulation runs, and does not provide researchers with information more immediately relevant on a practical level, i.e., how likely is their particular experiment to produce a desirable or an adverse outcome.

BP1s, and in particular CRM, might offer a sense of control and versatility and a shot at pinpointing the MTD early on by settling on it. But we have no good way of knowing whether the experiment really settled on the right dose, before it is too late. These designs necessitate intricate tweaking and calibration that Lee and Cheung described as a time-consuming search in multidimensional space \cite{LeeCheung09}. Even in recent years, some of the best-known CRM design experts still occasionally find themselves stumped by its unexpected behavior \cite[the latter is described in Supplement~B]{MathewEtAl04,RescheRigonEtAl08}. Last but not least: it is rather difficult to avoid perceiving the settling behavior as a sign of convergence, especially when accompanied with reassuring model-based terminology, and it is rather tempting to stop the experiment prematurely when settling occurs.

The nonparametric CCD offers MTD-selection performance and convergence properties similar to CRM's -- without any need for model skeletons, finely-tuned parameter distributions, or other sophisticated design tools. Only an interval and a set of dose levels are required. CCD experimental trajectories are somewhat quirky, tending to oscillate between temporary barriers and to occasionally generate ``incoherent'' transitions. If one chooses an interval design, care must be taken to communicate to clinicians that such oscillating behavior is likely. Besides that, CCD suffers from the same sensitivity to sampling order, and the associated $n^*$ variability common to all LMP1s. To complete the LMP1 picture, Supplement~E presents $n^*$ simulation distributions from a two-parameter BP1 design. The variability is just as high as with the one-parameter and nonparametric designs.

Using long memory or likelihood models for dose-finding can be a useful principle. Rather than the long memory itself, LMP1s' core vulnerability is their \textbf{``winner-take-all''} decision rule: attempting to allocate the $\widehat{MTD}$ itself at each step. After a handful of observations had accumulated, the incremental changes to the likelihood between successive cohorts are usually too small to force a dose transition (see the right-hand sides of Figures~\ref{fig:doughy00}-\ref{fig:N08}), and they continue to diminish as the experiment progresses. However, the $\hat{F}$ values underpinning the likelihood are still very imprecise at these small samples. Hence the ``settling'' phenomenon, with its poor predictive abilities with respect to MTD selection (see Section~\ref{sec:settl} and Figure~\ref{fig:StagEst}), and the resultant sensitivity to sampling order. Since BP1s weight the likelihood with a prior, the impact of new data is even smaller for them compared with non-Bayesian designs, settling is earlier and more pervasive, and outcomes become sensitive to both order and to predictive-prior weight (see Table~\ref{tbl:prior}, and also an experiment described in Supplement~B).

As briefly mentioned in Section~\ref{sec:prelim}, there exist Bayesian dose-finding approaches that attempt to optimize \emph{information collection} during dose assignments, rather than optimizing treatment. They are known as Bayesian Decision Procedures or BDP \cite{WhiteheadBrunier95,WhiteheadEtAl01}. Due to space limitations, and since these designs have been rarely discussed or put to the test, we have not included them in this article.

Interestingly, under a one-parameter model the optimization of treatment also optimizes information collection about $F^{-1}(p)$, and therefore at face value one-parameter CRM can be seen as both a BP1 and a BDP. This is a sleight-of-hand: proper Bayesian methods such as BDP assume that the model is a reasonable attempt to describe reality. By contrast, CRM's one-parameter approach is universally described as a ``working model'', rather than a realistic approximation of $F$. Researchers trying to construct a one-parameter CRM model around a realistic view of $F$ and to incorporate prior scientific knowledge into it, are quite likely to fall flat as happened to Neuenschwander \emph{et al.} \cite{Neunch08}. This is because such attempts inevitably abandon some crucial element of the delicate balance between operating characteristics, which is the real purpose towards which CRM has retooled Bayesian machinery.

This perspective leads to an interesting analogy between CRM and U\&D: both had originated from statistical intuition rather than from a well-established system of theoretical results, and both were later widely adopted for their ``side-effects". With U\&D the side effect is its Markov~chain behavior and the fast convergence to that behavior. With CRM, it is the ability to control early-cohort behavior, followed by the settling on a single dose.

Some interesting recent attempts to modify BP1 dose-assignment rules \cite{JiEtAl07,JiEtAl10,YinYuan09} will probably not resolve order sensitivity, unless the underlying loss function is modified to discourage a winner-take-all solution. Bartroff and Lai \cite{BartroffLai10} and Azriel \emph{et al.} \cite{AzrielEtAl11}, both writing about ``the treatment vs. experimentation dilemma'', each offered a new design. We have been able to examine the Azriel \emph{et al.} design, and it does not alleviate the variability in $n^*$ (see figure in Supplement~E).


This brings us to U\&D, a design family that has received scant attention in recent Phase~I literature. Regardless, U\&D is used every day in dozens of similar applications, including dose-finding anesthesiology experiments \cite{PaceStylianou07}. Common wisdom in Phase~I methodological circles, as indicated by the tone of article introductions, has been that LMP1s, in particular BP1s, deliver far better MTD-selection success rates than any other approach including U\&D. In view of Section~\ref{sec:numer}, this notion has been misguided. One can trace its origins to studies that inadvertently used the last assigned dose to represent U\&D's ``estimate'' \cite{OQuigleyChevret91}, or to studies that chose scenarios highly favorable to the BP1 models examined. But most often, LMP1s have been compared only to each other or to `3+3', whose MTD-selection performance is poor.

U\&D, as well, has its limitations, such as a relatively rigid design and no ``natural'' standard estimator. The latter problem has been largely mitigated with the adaptation of isotonic regression to U\&D \cite[Section~3.3]{StylianouFlournoy02,Oron07}. If statisticians invest more resources in nonparametric dose-finding estimators or in U\&D, this design's performance is likely to improve even further.

The well-known ``two-stage'' LMP1 approach \cite{Storer89,Storer01,IasonosEtAl08} starts with a stage of single-patient cohorts, escalating until the first toxicity. This stage's dose-transition rules are identical to a median-targeting U\&D \cite{DixonMood48}. However, the first-toxicity transition to LMP1 would occur too early to avoid the side effects described in this article.  A simultaneous hybrid U\&D-LMP1 approach with interesting properties was developed by Narayana in the 1950s, and recently rediscovered and discussed by Ivanova \emph{et al.} \cite{IvanovaEtAl03}. A newer hybrid design incorporating U\&D in a role analogous to the sequential probability ratio test's \emph{``continue sampling''} option \cite{Wald45} was presented in ref. \cite[ Ch.~5]{Oron07}. It succeeds in increasing $\overline{n^*}$ compared with U\&D, while retaining low variability and somewhat improving MTD-selection performance (see Supplement~E). The Narayana design can be seen as a simple special case of this hybrid design family. This is an area of ongoing research.

Beyond debating and improving designs, there are simpler recommendations methodologists can agree upon. First and foremost, defining a Phase~I study as an exercise in \emph{dose-selection} rather than  \emph{dose-estimation} further degrades the amount of information, in an application whose main challenge is information scarcity. It will be beneficial for the field if we succeed in convincing clinicians to accept an \emph{estimate} of $F^{-1}(p)$ on a continuous scale, rather than try to pinpoint the best candidate from a fixed dose set. This is especially true when there is no good MTD candidate to be found, or when two dose levels are nearly equally suitable. It should be pointed out that the MTD-selection success rates presented in Section~\ref{sec:numer} are somewhat optimistic, since the simulations excluded scenarios belonging to these two categories. Moving from a selection paradigm to an estimation paradigm will also remove some of the practical constraints that make Phase~I designs so cumbersome. This web of constraints might have contributed to the emergence of LMP1s, an approach that makes the unrealistic promise of solving the estimation, selection and treatment requirements all at once with a small sample of binary observations.

Another salient example for the MTD-selection paradigm's undesirable effects is the misrepresentation of experimental outcomes. The sharply different performance outcomes for $l=7$ and $l=4$ in Section~\ref{sec:rand} do not represent the true difference in \textbf{the information gathered about $F^{-1}(p)$.} Under both settings we have 25 binary data points to make the final estimate, and with $l=7$, the information is generally better concentrated around $F^{-1}(p)$. Moreover, toxicities can be better controlled with a finer dose spacing, and treating patients one level off the MTD is not as detrimental. Therefore, moving from selection to estimation will allow for the use of finer dose scales, almost regardless of sample size. To make sure the entire dose range is accessible during the experiment, the initial run until the first toxicity can be carried out on a coarser dose set (e.g., skipping one dose level at a time).

As long as the experiment's goal is defined as dose selection rather than estimation, methodologists should highlight the tradeoff between $n$ and $l$ (more patients whenever possible, less dose levels to improve MTD detectability), and align expectations based on the final choice of $n$ and $l$.  For example, with $n\approx 20$ and $l\geq 6$, one should not expect better than even odds of finding the true MTD under any design. As long as the dose-selection paradigm dominates, methodologists should strive to make sure there are at least $5-6$ patients per dose level to ensure reasonable prospects for MTD-selection success.

\bibliographystyle{plain}
\bibliography{phdplus}


\newpage
\Roman{figure}
\onehalfspacing
\begin{Large}
\begin{center}
{\bf Supplement for \\ ``Small-Sample Behavior of Novel Phase~I Cancer Designs''\\} 
by Oron and Hoff
\end{center}
\end{Large}
\medskip

\section*{A. The Chevret One-Parameter CRM Model}

Most current and published CRM experiments have used the ``power'' model described in the article's Section~2.1. However, some studies such as \cite{FlinnEtAl00} (used for the article's Fig.~1) and \cite{DoughertyEtAl00} (Fig.~2) use a more sophisticated version developed by Chevret \cite{Chevret93}. It is sometimes misunderstood as a one-parameter logistic model with the location parameter fixed. In fact it uses a one-parameter logistic ``skeleton'', somewhat re-parametrized -- and then transforms it \emph{horizontally}, i.e. on the dose scale. The skeleton parametrization is

\begin{equation}\label{eq:chevret0}
\Gamma(\xi)=1+\exp\left[\beta_0-\theta\xi\right].
\end{equation}

The data-estimable parameter $\theta$ affects both location and scale. However, the curve is logistic only when plotted vs. the transformed doses $\xi_u$, which are related to the original dose levels $d_u$ via

\begin{equation}\label{eq:chevret1}
\Gamma(\xi_u ; \theta_0)=\phi_u,\ \ u=1,\ldots,l,
\end{equation}

where $\phi_u$ is initial toxicity-rate estimate at $d_u$ according to researchers' prior knowledge, as in the ordinary ``power'' model, and $\theta_0$ is the prior mean of $\theta$. Thus, the $\xi_u$ are found by back-calculation. Ostensibly this allows for the same flexibility as with the ``power'' model while maintaining coherent curve that is often used in literature. However, the lateral transformation makes it hard to envision the final dose-toxicity curve.

\section*{B. Description of Several Additional Experiments}
\begin{figure}
\begin{center}
\includegraphics[scale=0.4]{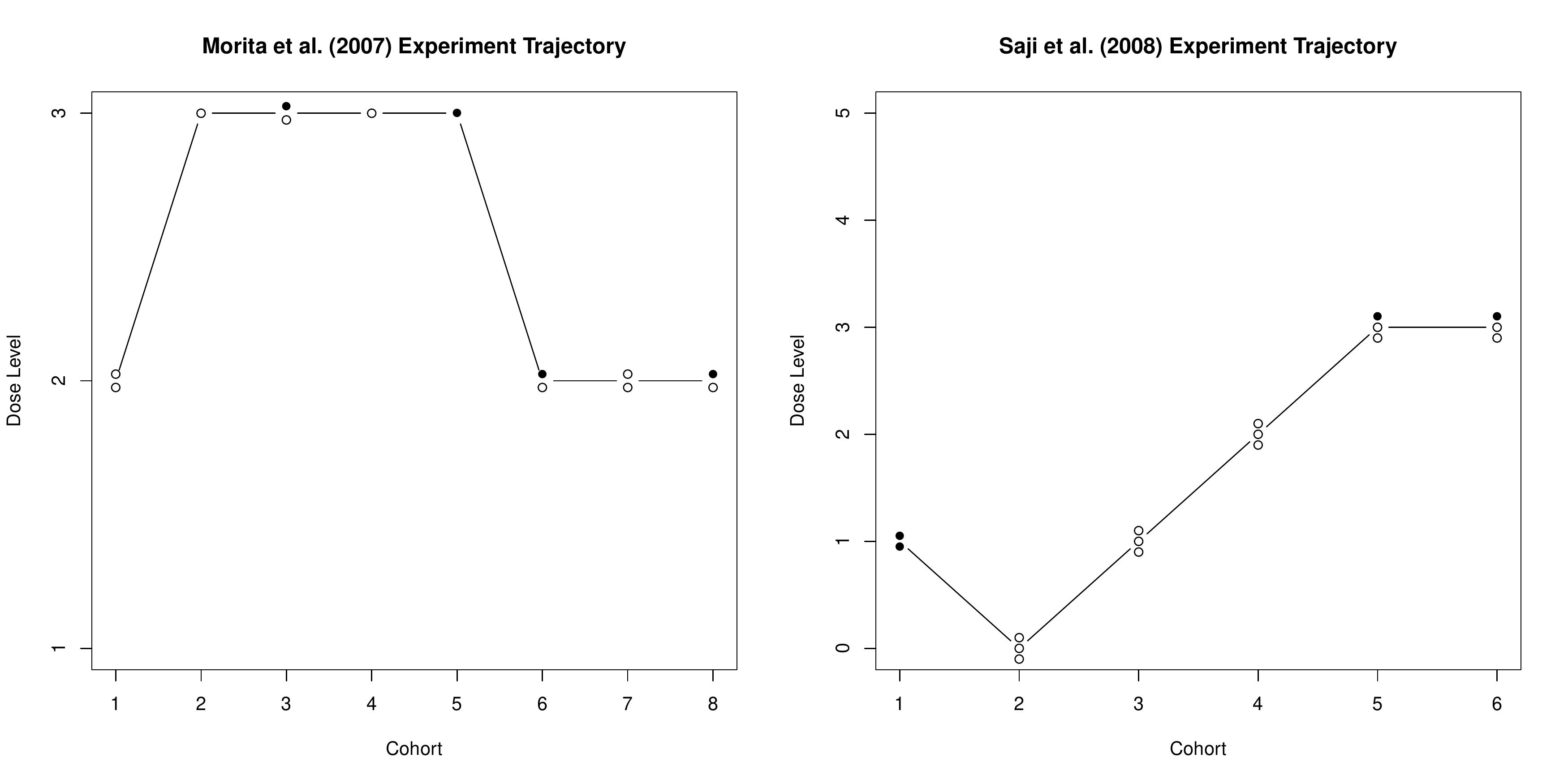}
\end{center}
\caption{Trajectories of the \cite{MoritaEtAl07} and \cite{SajiEtAl07} experiments, using the conventions from Figure 1 of the main article.}\label{fig:moritasaji}
\end{figure}

Fig. \ref{fig:moritasaji} shows the trajectories of two recent Japanese CRM studies. Morita et al. \cite[left]{MoritaEtAl07}  targeted $p=0.2$, obeyed the Goodman et al. \cite{GoodmanEtAl95} constraint, and had three levels with $x_1=d_2$. To control toxicities, most cohorts at $d_3$ were limited to single patients, compared with two patients per cohort at lower levels. After no DLT's on cohort~1 the dose was escalated, with the first DLT observed on the third patient at $d_3$ (fifth from the start). However, CRM (which used the consensus of four ``power'' models with different skeletons, perhaps a precursor to Yin and Yuan's \cite{YinYuan09} Bayesian Model Averaging work) still prescribed a repeat of $d_3$, which produced one more DLT-free patient followed by another DLT. The final six patients were treated at $d_2$, with $2$ DLT's. The experiment ended after $13$ patients due to ``settling'', recommending $d_2$. Even though the observed DLT rate eventually exceeded the target rate $p$ on both levels $2$ and $3$ ($2/8$ and $2/5$, respectively), $d_1$ was never allocated during the experiment. Another recent CRM study from Japan \cite[right]{SajiEtAl07} is shown here not for its model properties, but because the sequential allocation to all of this trial's $6$ cohorts would have been completely identical, had the researchers used the '3+3' protocol instead.

In France, Resche-Rigon et al. \cite{RescheRigonEtAl08} implemented CRM in a Phase~II pediatric experiment targeting a $p=0.1$ treatment-failure rate, with cohort size 1. There were six levels using Chevret's model and the skeleton $(0.001, 0.05, 0.1, 0.3, 0.6, 0.7)$.\footnote{Technically, this was a dose de-escalation trial. The probabilities are failure probabilities and $d_1$ was in fact the highest dose. But both the trial description in the article and our presentation here reverse the outcomes, so that the terminology is the same as that of a toxicity escalation trial.} Thus, the prior $\widehat{MTD}$ is $d_3$, which is where the trial began.

The very first outcome was a failure, pushing the dose from $d_3$ down to $d_1$. Allocations remained at $d_1$ for the remainder of the experiment, despite 0-of-10 failures observed. Calculations indicated that only after 0-of-14 at $d_1$, escalation to $d_2$ would have been finally allowed. The authors concluded that the first observation was ``inordinately influential.'' For future studies, the researchers suggested imposing an ad-hoc weighting scheme on the likelihood calculations, discounting the impact of observations as the recede into the experiment's past. This turns the design into a compromise between long and short memory, albeit with unclear properties.

A point not noticed by the authors is that the wait at $d_1$ has been extended by the extremely low prior skeleton value for this dose. This value places it far out on the low tail of the transformed logistic curve (see the explanation of Chevret's model in Section~A above). Hence, successes observed at $d_1$ have a very weak impact upon the estimates at the other doses. Had the skeleton value for $d_1$ been $0.01$ instead of $0.001$, only $8$ successes at $d_1$ would have been needed before the next dose transition, rather than $14$. This again demonstrates the intricate interplay between CRM design parameters, and its behavior in practice. The senior author of this study is a well-respected CRM expert who has contributed substantially to Phase~I methodology and standards. It should also be mentioned that the $k$-in-a-row U\&D design targeting $p\approx 0.109$ necessitates exactly six (6) consecutive successes before each dose escalation, regardless of order in the experiment or previous outcomes.

In a more recent Phase~II dose de-escalation efficacy study with 5 levels, also targeting $p=0.1$ failure rate, the same authors decided not to use the past-discounting scheme they had developed \cite{ZoharEtAl12}. Two failures at patients number 6 and 10 pushed the experiment up to the $d_4-d_5$ range for the next 10 patients. When the 25-patient trial was over, estimates suggested that either $d_3$ (6 patients treated) or $d_2$ (only 1 patient treated) is the MED, depending upon the estimation method.

\section*{C. Similarities and Differences between `3+3' and Up-and-Down}
Even though we have not found definite historical proof, is quite likely that the `3+3' protocol is inspired by group Up-and-Down (GU\&D) designs. These designs has been in use since the 1960's \cite{Tsutakawa67}, and this is also the time frame when `3+3' begins to make its appearance. In GU\&D designs, a cohort of $k$ subjects is treated simultaneously. If $b$ or more toxicities are observed, treatment descends one level down. If $a$ or less are observed, it escalates one level up. Otherwise, the next cohort receives the same treatment. Obviously, $0\leq a<b\leq k$. Choice of stopping rules and estimation method is left up to the researchers' (and consultants') discretion. Gezmu and Flournoy \cite{GezmuFlournoy06} introduce the useful shorthand terminology GU\&D$_{(k,a,b)}$ to describe any design of this family.

The version of `3+3' most commonly quoted nowadays run as follows \cite{RosenbergerHaines02}:
\begin{enumerate}
\item Start at the lowest (or sometimes second-lowest) level.
\item Treat cohorts of $3$ subjects at a time.
\item If this is the first cohort at the present level do as follows: if no toxicities are observed, escalate; if $2$ or $3$ are observed, descend; if $1$, treat another cohort at the same level.
\item If this is the second cohort at the present level, consider all $6$ subjects. If $2$ or more toxicities were observed, descend. Otherwise escalate.
\item If a third cohort is mandated for any given level, then the experiment stops.
\item The MTD estimate is the highest level $d_u$ such that $\hat{F}_u<1/3$.
\end{enumerate}

Some variants use even more aggressive stopping rules, such as stopping the experiment after encountering a level with $2$ toxicities out of $6$ and declaring the next-lowest level the MTD, or (similarly) declaring a level with $1$ of $6$ to be the MTD.

The beginning of a `3+3' experiment looks just like a GU\&D$_{(3,0,2)}$ (which, by the way, targets $Q_{0.347}$). However, upon visiting the same level a second time, the next transition decision is changed to something like a GU\&D$_{(6,0,2)}$ transition (targeting $Q_{0.181}$) -- with the important distinction that in a GU\&D experiment the decision is based only upon the current cohort and not upon less recent ones (a genuine GU\&D$_{(6,0,2)}$ would treat all $6$ subjects at once). Furthermore, each time a 3+3 experiment visits a new level, decision rules revert to the GU\&D$_{(3,0,2)}$-like stage.

In summary, these are the major differences between `3+3' and the U\&D family:

\begin{enumerate}
\item The '3+3' design switches mid-experiment back and forth between $1$-cohort and $2$-cohort transition rules.
\item The 3+3 two-cohort rule does not necessarily involve the $2$ most recent cohorts.
\item The two rules (when used each exclusively) target different percentiles.
\item Even more importantly, '3+3' has aggressive stopping rules prohibiting the administration of any single dose to more than $2$ cohorts; U\&D designs have no such constraint.
\item These differences combine to spoil random-walk properties. Unlike U\&D, one cannot describe the trajectory of a `3+3' as a simple Markovian random walk with tractable asymptotic behavior (even though `3+3' is still a stochastic design of sorts).
\item Last but not least, the '3+3' MTD estimate is usually the stopping dose or the one below it. With U\&D designs, the estimate is not related to the last administered dose, but is instead calculated using information gathered from all the experiment's trials, via some averaging scheme or isotonic-regression interpolation \cite{StylianouFlournoy02,Oron07}.
\end{enumerate}

\section*{D. Supplementary Simulation Information}
\subsection*{Model Curves and Simulated Scenarios for Sec. 4.1-4.3}
Rather than use arbitrarily chosen, rounded toxicity values at the dose levels (as is often done in BP1 simulation), we preferred to simulate $F$ using standard distributions, which approximate scenarios that can be realistically encountered in practice, and which are commonly used to model dose-response dependence. Curve families used include Logistic, Normal, Gamma, Weibull, Lognormal and uniform. Dose levels were always uniformly spaced. The CRM model details were provided in the article body.

\begin{figure}
\begin{center}
\includegraphics[scale=0.7,angle=90]{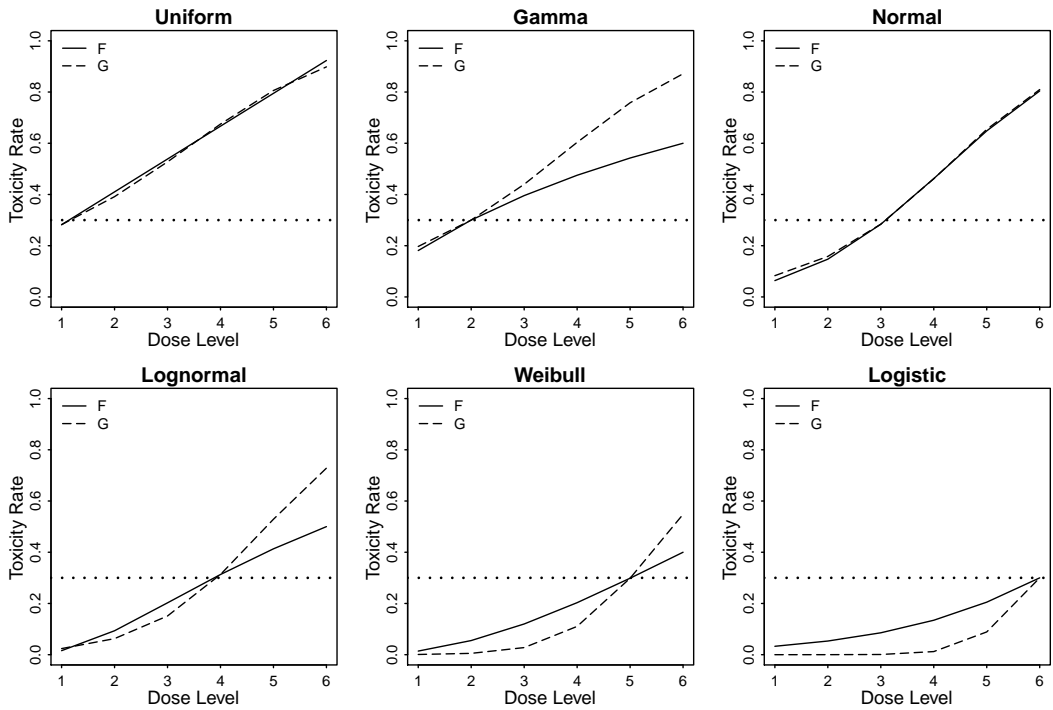}
\caption{Toxicity curves (solid lines) for the six scenarios. The dashed lines show the CRM `power' model curves that match each toxicity curve exactly at the MTD. MTDs are levels 1 through 6, increasing from left to right and top to bottom. The target toxicity rate itself ($p=0.3$) is indicated via horizonal dotted lines.}\label{fig:models}
\end{center}
\end{figure}

Figure \ref{fig:models} shows toxicity curves from the simulation setup having $l=6$ uniformly-spaced levels, and $16$ cohorts of $k=2$ subjects each per run. The figure also shows the model curve that matches $F$ exactly at the MTD. We chose $6$ scenarios that are sufficiently different, realistic, present different levels of challenge to the CRM designs, and also have different levels as MTDs. As specified in the article, the one-parameter ``skeleton'' used was $\phi=\left(0.05,0.11,0.22,0.40,0.60,0.78\right)$, which is equivalent to a logistic curve with location parameter $\mu=0.75$ and scale parameter $\sigma=0.2$, if we assign to the dose levels the evenly spaced numerical values $\left\{1/6,1/3,\ldots ,1\right\}$. This skeleton closely resembles the convex ones preferred by many CRM researchers. The prior distribution on the single data-estimated `power' parameter was Lognormal. Prior A (the main one used to produce all figures) had the Lognormal parameters $\mu=-0.2,\sigma=0.85$. Prior~B that placed more weight on higher level, pushing the runs more aggressively upward, used $\mu=0.0,\sigma=\sqrt{1.34}$. This, by the way, is the ``default'' prior in the \texttt{crm} R function by Cheung. It is preferred by researchers who use convex skeletons, because on these skeletons it tends to produce a nearly-uniform predictive prior for the MTD. Prior~C reflects a strong belief that the MTD is \emph{not} at the higher doses. It uses $\mu=-0.5,\sigma=0.6$.

As can be seen, the Uniform and Normal scenarios are matched very closely by the model, which indeed meets the restrictive Shen-O'Quigley \cite{ShenOQuigley96} convergence criteria for these scenarios. The Gamma and Lognormal scenarios only meet the relaxed criteria suggested by Cheung and Chappell \cite{CheungChappell02}, indicating somewhat slower convergence; while the Weibull and Logistic scenarios are not guaranteed to converge to the MTD.

\subsection*{Details of the Random $F$ Simulation, Sec. 4.5}

As stated in Section~4.5, in our view the current dominant practice in Phase~I performance simulations is flawed. The few consciously chosen scenarios for $F$ on which performance statistics are reported, are typically more favorable to each author's preferred method. Moreover, such simulations under-represent the range of variability encountered in practice, and often some of the scenarios are unrealistic or uninteresting.

An alternative paradigm is that of randomly-generated scenarios, pioneered by O'Quigley et al. \cite{OQuigleyEtAl02} and Paoletti et al. \cite{PaolettiEtAl04}. The rationale is that scenarios are simply $l$-tuples of positive increasing numbers in $[0,1]$. In other words, the space of all possible scenarios is the $l$-dimensional unit simplex. One can then craft strategies for random sampling from that simplex.

Besides the above-cited authors, Ivanova et al. \cite{IvanovaEtAl07} and Azriel \cite{Azriel12}, as well as ourselves, have also attempted such sampling, each using a different paradigm. We have chosen to use random Dirichlet vectors, as introduced elsewhere \cite{OronEtAl11ccd}. In each scenario, a set of $l$ increments (including the increments between $0$ and $F\left(d_1\right)$) was simulated as a single multivariate Dirichlet random variable. The value of $F$ at $d_u$ in scenario $j$ can be written as

\begin{equation}
F^{(j)}\left(d_u\right)=\sum_{m=1}^u X^{(j)}_m,\ \ \ \mathbf{X}^{(j)}\sim Dirichlet\left(\alpha_1,\ldots \alpha_{l}\right).
\end{equation}

The parameters $\left(\alpha_1,\ldots \alpha_{l}\right)$ were not directly chosen, but were randomly and independently drawn for each scenario. This already assures two layers of randomization between the user and the outcome, beside the randomization of toxicity thresholds for each simulation run.

Our experience with scenario randomization has emphasized the importance of calibrating parameters to ensure scenarios will be realistic and interesting. For example, the $4$-tuple $(0.0001,0.001,0.01,0.1)$ is just as valid a sample as $(0.1,0.3,0.5,0.6)$. However, when targeting a toxicity rate of $0.3$, the former would generate incredibly boring runs, most likely ending up at the top dose with zero overall toxicities -- regardless of design. Rather than design properties, such a scenario exhibits an awfully misinformed choice of $F$.

To make realistic scenarios more likely than non-realistic ones, we added a few more layers of randomization. The Dirichlet vectors are the sum of a baseline drawn from a uniform distribution, and a Gaussian ``bump'' of random magnitude, location and spread, indicating the threshold-population's main mode. Even before that, the vector is lengthened by a ``padding'' of random length. The $l$-tuple is then a random sample from a longer vector. This enables the mode to be above or below the dose range, generating concave or convex $F$ forms aside from sigmoid ones. The vector's random length also allows for a range of spacing magnitudes between $F$'s values (i.e., steeper or shallower curves). A further twist is added by optionally raising all the $F$ values to a random power in $[p,1/p]$, and optionally adding a small random constant to all the values (constrained to ensure they still all fall in $[0,1]$).

In addition, we keep in mind that the main performance metric is the binary one of finding the true MTD (rather than a continuous measure such as the estimate's RMSE). Hence, it might be unfair and misleading to include scenarios in which the MTD is ambiguous, or in which even the MTD is a rather poor choice because its $F$ value is very far from target. Therefore, any candidate $l$-tuple was vetted to make sure it meets these conditions. Specifically, for $l=7$ we included only scenarios for which $F$ at the MTD was in $[0.22,0.38]$, it was the only one permitted inside the interval, and all other levels had $F$ values at least $0.06$ further away from $0.3$, compared with the MTD's exact $F$ value. For $l=4$ the respective cutoffs were $[0.18,0.42]$ and $0.09$. Scenarios in which some $F$ increments were too large or too small, were also excluded. As a result, the ensembles are somewhat more well-behaved than the originally generated ensemble distribution. 

Finally, in order to ensure a reasonably uniform MTD distribution, we generated a super-ensemble of $\gg M$ scenarios, then performed a stratified random sample of size $M$ from that super-ensemble. For $l=7$, the final simulation ensemble had a true MTD distribution of $320$ runs per level for $d_2$ through $d_6$, and $200$ runs each for the boundary levels. For $l=4$, the MTD distribution was $(400,600,600,400)$.

\begin{figure}
\begin{center}
\includegraphics[scale=0.8]{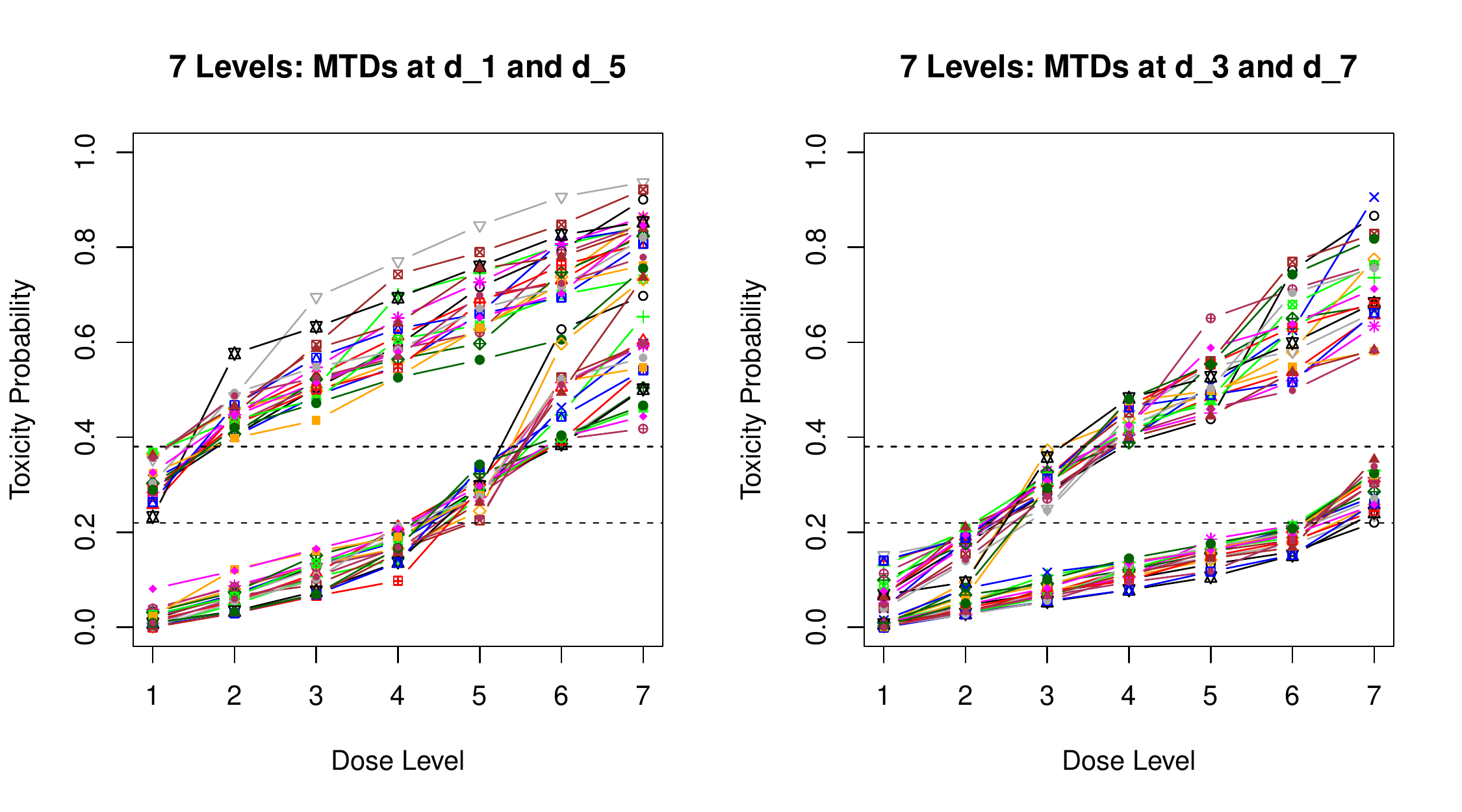}
\caption{Toxicity curves for four samples of size 20 out of the 2000-scenario random-$F$ simulation for $l=7$. Shown are scenarios with the MTD at $d_1$ and $d_5$ (left), and $d_3$ and $d_7$ (right). The dashed horizontal lines indicate the interval in which the MTD's $F$ values must fall.}\label{fig:random7}
\end{center}
\end{figure}

The scenario generation code is enclosed in Supplement~F below. Various tuning parameters help control the magnitude of the different randomization components, and the values we used are listed as code comments. Figure~\ref{fig:random7} shows samples of size 20 from the generated scenarios for the various MTD levels. The target rate's asymmetry (closer to 0 than to 1) causes scenarios with a high MTD dose to be far less diverse than those with a low MTD. The largest variability in both form and $F$ values at the doses, is observed among scenarios whose MTD is $d_3$ (right).

The CRM design skeletons for this simulation were, as described in the text, the ones from \cite{FlinnEtAl00} and \cite{PistersEtAl04} for $l=7$ and $l=4$, respectively. For the latter, the prior distribution for $\theta$ was the published one, i.e., log-Normal with $(\mu,\sigma^2)=(0,1.8)$. For the former, the design originally used the \cite{Chevret93} model and an Exponential prior. To make the comparison between $l=7$ and $l=4$ more equitable, we used the ``power'' model instead, with the \texttt{`dfcrm'} default log-Normal$(0,1.34)$ prior. That package was also used to calculate the dose allocation decisions.

\section*{E. Run-to-Run $n^*$ Variability of Some Additional Designs}

\begin{figure}
\begin{center}
\includegraphics[scale=0.6]{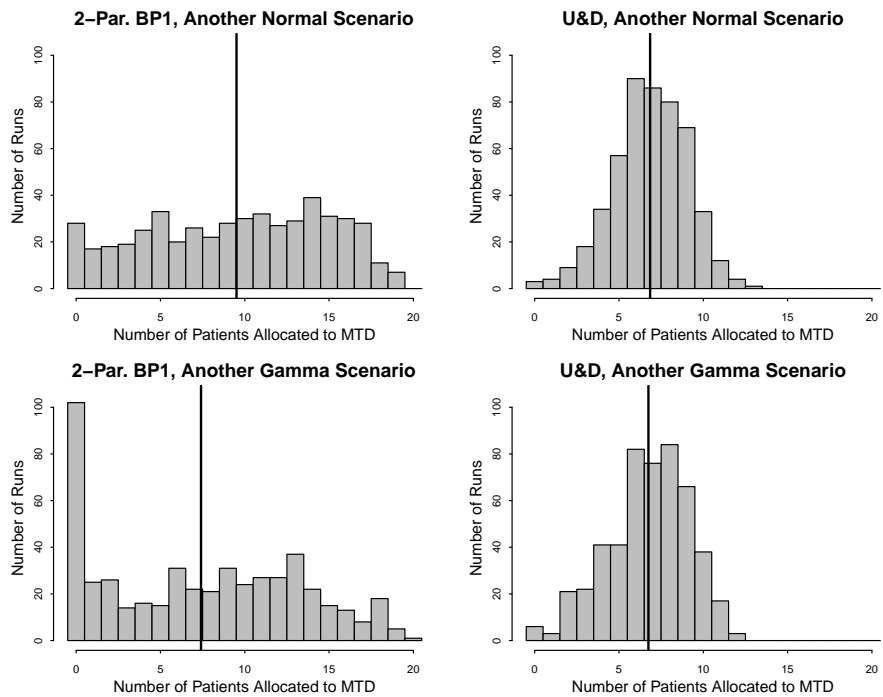}
\end{center}
\caption{A plot analogous to the main article's Figure~5 (distribution of $n^*$), but from a different simulation run that included a 2-parameter logistic BP1 (left) and a $k$-in-a-row U\&D design(right). Simulation details are in the text.}\label{fig:BP1}
\end{figure}

The article makes heavy use of one-parameter CRM experiments and examples, because CRM is by far the most well-known and implemented LMP1 design. However, the order sensitivity and $n^*$ variability are universal LMP1 features. This is exemplified in the article via CCD examples. Figure~\ref{fig:BP1} shows examples from a simulation that included a two-parameter BP1 using a location-scale logistic model. That simulation had 500-run ensembles with a cohort size of 1. There were $l=8$ dose levels, and the U\&D design used was ``$k$-in-a-row'', as in the article's random-$F$ simulations.

Shown are the $n^*$ distributions of the first 20 allocations of 2-parameter BP1 (left) and U\&D (right), under two scenarios. These are not the same ``Normal'' and ``Gamma'' scenarios as in the main article's Section~4. Both MTDs are fairly easy to detect in terms of dose spacing. However, the ``Gamma'' MTD's 2-parameter prior-predictive weight was smaller than the ``Normal''. As suggested in the article (Section 5.1), the more parameters in the curve, the more influential the prior might become. Thus, the 2-parameter runs settle earlier than 1-parameter CRM runs (5-cohort settling observed by the 10th allocation in $45\%$ of Normal runs and $42\%$ of Gamma runs, compared with $33\%$ and $32\%$, respectively, with one parameter CRM). In any case, the two-parameter BP1 variability in $n^*$ is strikingly similar to that observed with one-parameter designs, while U\&D is robust by comparison, both between runs and between scenarios.

\begin{figure}
\begin{center}
\includegraphics[scale=0.7]{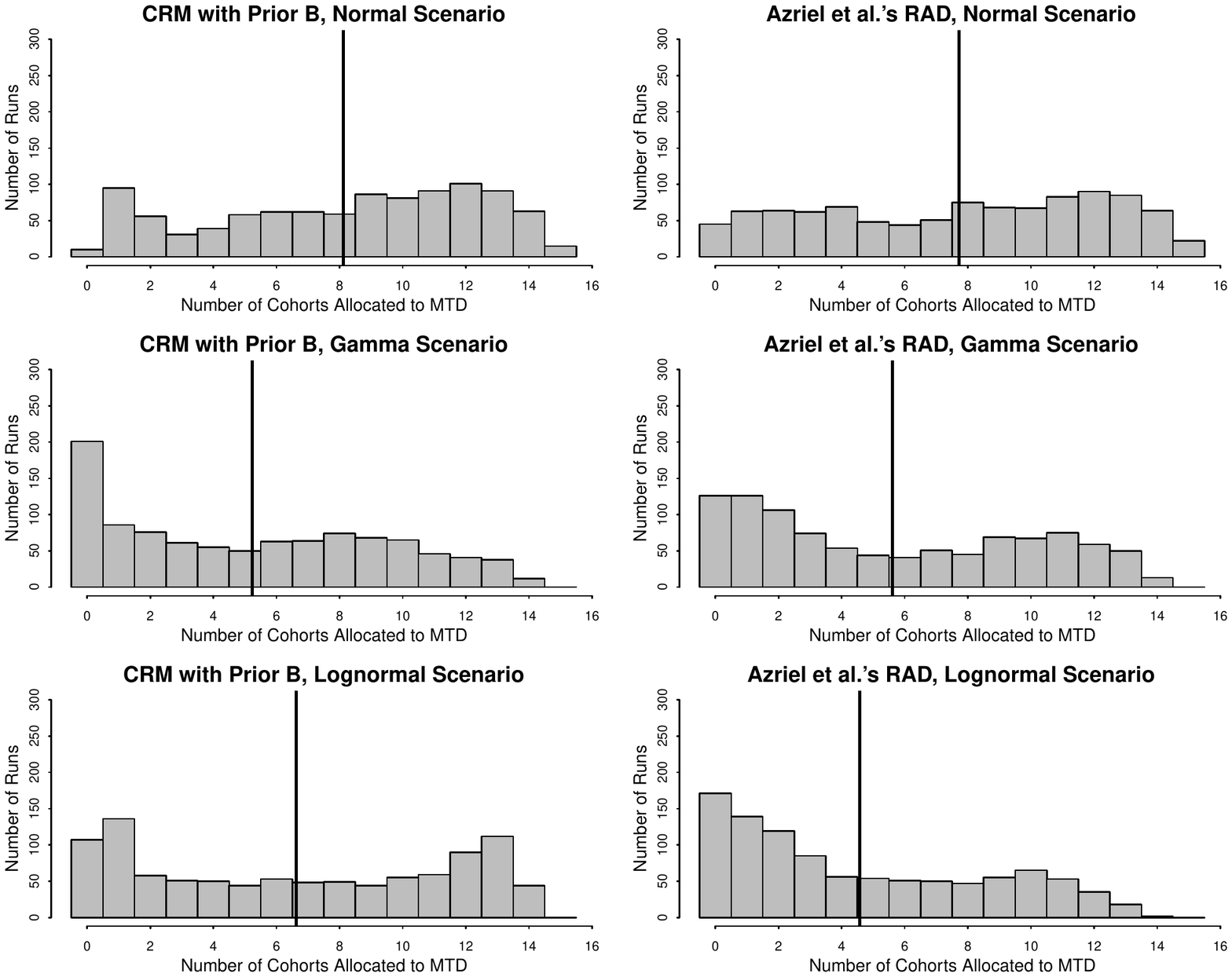}
\end{center}
\caption{A plot identical to the main article's Figure~5 (distribution of $n^*$), except that here the two compared designs are CRM with Prior~B, the ``default'' prior (left), and Azriel et al.'s RAD (right).}\label{fig:priorB}
\end{figure}

Figure~\ref{fig:priorB} returns to the simulation design used in the article. On the left is the same one-parameter ``power'' model from the article, but with the ``default prior'' (Prior~B). The variability in $n^*$ is practically identical to that observed in the article's Figure~5.

The right-hand-side of Figure~\ref{fig:priorB} shows the $n^*$ distributions on the same scenarios, for Azriel et al.'s \cite{AzrielEtAl11} Random Allocation Design (RAD). RAD takes a nonparametric LMP1 known as ``isotonic regression design'' \cite{LeungWang01}, and adds randomization. The original design is closely related to CRM, allocating to the level whose isotonic-regression $\hat{F}$ estimate is closest to $p$. Under RAD, the next cohort might be assigned instead to the dose on the opposite side of target, according to a random draw whose probability is inverse to $n$. While the ``isotonic regression design'' does not converge, Azriel et al. proved that RAD does converge in probability to the MTD (but not almost surely). However, as Figure~\ref{fig:priorB} (right) shows, RAD behaves very poorly in terms of $n^*$ variability. Its average estimation performance for small samples is also unimpressive (data not shown). This suggests that while a simple weakening of the ``winner-take-all'' rule can lead to better convergence, a more careful modification than blind randomization is needed in order to improve small-sample behavior.

\begin{figure}
\begin{center}
\includegraphics[scale=0.7]{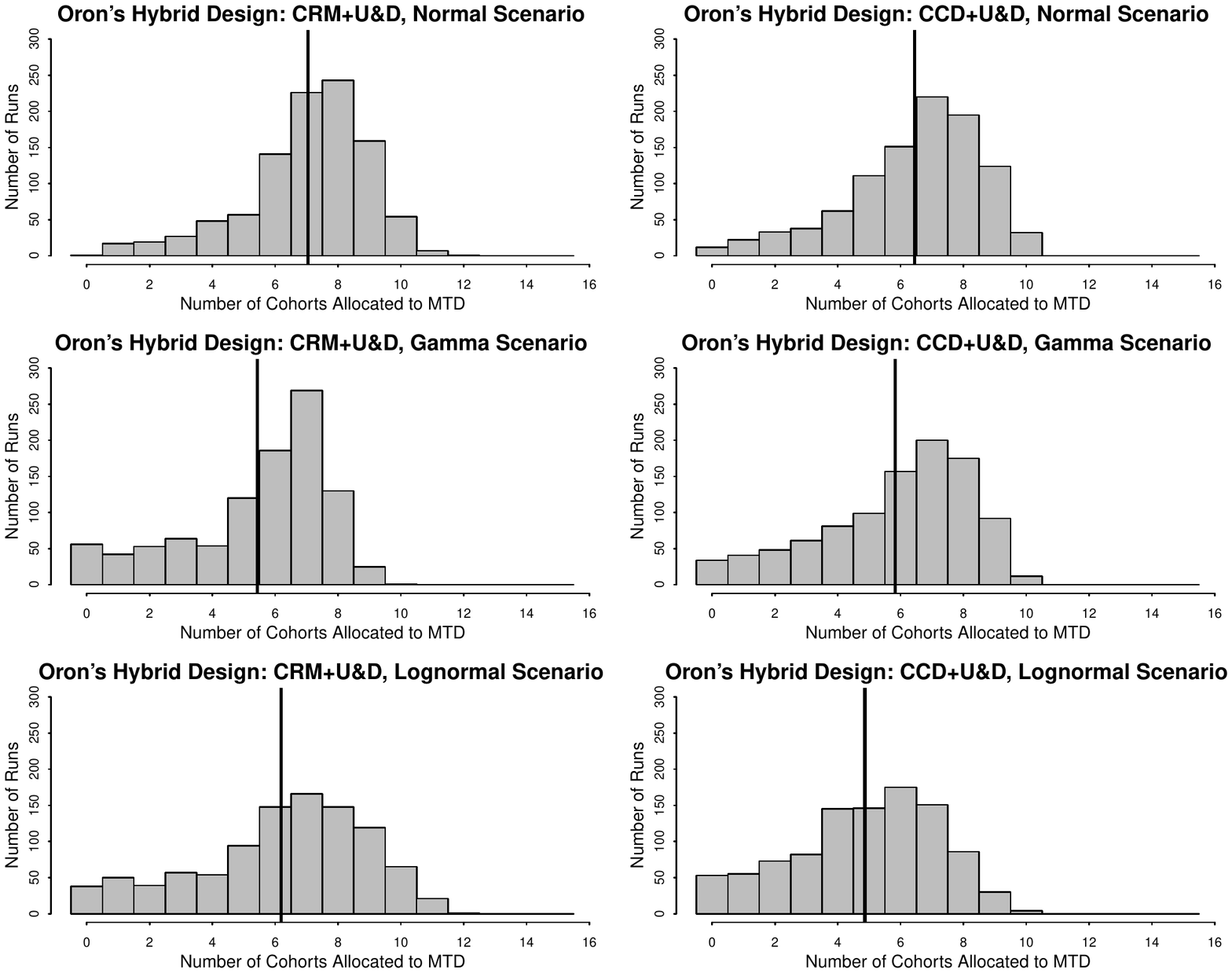}
\end{center}
\caption{A plot identical to the main article's Figure~5 (distribution of $n^*$), but with Oron's \cite{Oron07} hybrid design, combining U\&D with CRM (left) and CCD (right).}\label{fig:BUD}
\end{figure}

One such modification \cite[Ch.~5]{Oron07} combines U\&D with an LMP1 design. Unlike RAD's randomization, which inevitably leads to ``non-coherent'' assignment decisions (i.e., escalation following DLTs and vice versa), here the non-LMP1 rule is U\&D, which is in fact the design's default. LMP1 can override the U\&D assignment, only if the override passes a test of confidence. For example, if CRM indicates staying at $d_u$, while U\&D indicates escalation to $d_{u+1}$, the experiment will escalate {\bf unless} the combined MTD-predictive-posterior weight of all levels above $d_u$ is less than a fixed confidence threshold $\beta:\ 0<\beta<0.5$. For non-Bayesian designs, probability calculations or $p$-values are used instead of the posterior. Generally, early in the experiment the U\&D rule will be used exclusively, and gradually more LMP1 decisions will be accepted. Lower values of $\beta$ are conservative, while values close to $0.5$ are aggressive.

Figure~\ref{fig:BUD} and Table~\ref{tbl:BUD} show results of a U\&D-CRM combination with $\beta=0.25$, and a U\&D-CCD combination with $\beta=0.35$. The overall distribution of $n^*$ using this design remains similar to U\&D's (compare with Fig.~5, main article), but its center is shifted some 1-2 cohorts to the right. Average performance -- especially after 32 patients -- is also somewhat improved and less variable between scenarios, compared with CRM, CCD or U\&D alone. Unlike the tables in the article, in this table the best performance in each line item was bolded regardless of margin.

\begin{table}
\caption{Bulk performance of two hybrid designs (U\&D+CRM and U\&D+CCD), compared with the best of the main article's Table~1 (Section 4.3). For each of six scenarios, compared are the proportion of runs in which the correct MTD was selected, after $8$ (left) and $16$ (right) cohorts, respectively. The hybrid design is estimated using centered isotonic regression.}\label{tbl:BUD}
\begin{tabular}{p{1cm}cccccc}
\\
\toprule
 & \multicolumn{3}{c}{\bf After 8 Cohorts} & \multicolumn{3}{c}{\bf After 16 Cohorts} \\
Scenario &  U\&D+CRM & U\&D+CCD & Tbl. 1 Best &  U\&D+CRM & U\&D+CCD & Tbl. 1 Best\\
\midrule
``Uniform''  & 53.6 & 55.7 & {\bf 57.1} & {\bf 64.7} & 63.0 & 64.1  \\
``Gamma''  & 43.8 & 41.8 & {\bf 44.2} & 51.5 & 51.5 &{\bf 53.2}  \\
``Normal''  & 56.9 & 55.5 & {\bf 57.8} & {\bf 68.6} & 67.3 & 67.5  \\
``Lognormal''  & 44.5 & 36.1 & {\bf 46.7} & 55.5 & 48.5 & {\bf 59.3} \\
``Weibull''  & 36.0 & 35.1 & {\bf 39.0} & 46.1 & 46.3 & {\bf 47.2} \\
``Logistic''  & 23.7& 25.4 & {\bf 32.2} &    34.9 & 51.0 & {\bf 54.6} \\
\bottomrule
\end{tabular}
\end{table}

\section*{F. Code}

\subsection*{Centered Isotonic Regression}

\begin{small}
\singlespacing
\begin{verbatim}
cirPAVA <-function (y,x, wt=rep(1,length(x)),boundary=2,full=FALSE,dec=FALSE,wt.overwrite=TRUE,
	xbounds=c(0,1),ybounds=c(0,1))
{
# Returns centered-isotonic-regression y values at original x points
# Assaf Oron, 10/2007 (some modifications 7/2012)
#
### ARGUMENTS:
# y:	 y values (responses). Can be a vector or a yes-no table (for binary responses)
#    it is Given as first argument, both for compatibility with 'pava' and to enable parallel running via 'apply' type routines
# x: 	treatments. Need to be pre-sorted in increasing order, with order matching y's
# wt: weights. Will be overwritten in case of a yes-no input for y
# boundary: action on boundaries. Defaults to 2, which is analogous to 'rule=2' on function 'approx', i.e.
# 	returned y values are constant outside the boundaries.
# 	boundary=1 does linear extrapolation. In addition, one can impose boundaries as inputs or
#	 augment the output with boundaries, as discussed in  the dissertation text.
# full: logical. If FALSE, only point estimates at x values are returned
# dec: Whether the true function is assumed to be decreasing. Defaults to FALSE
# wt.overwrite: whether, in case of yes-no input, the weights should be recalculated as row
# observation counts
# xbounds,ybounds: numeric vectors of length 2. Boundary conditions imposed on the extrapolation.
# 	Only relevant if `boundary==1'.
### adapting input in case of yes-no table, and some validation

ll=dim(y)
if (length(ll)>2) stop ("y values can only be a vector or yes-no table.")
if (length(ll)==2 && ll[2]>2) stop ("y values can only be a vector or yes-no table.")

if (length(ll)==2 && ll[2]==2) { # converting a yes-no table to x-y

    n.u<-y[,1]+y[,2]
    x<-x[n.u>0]
    y<-y[n.u>0,1]/n.u[n.u>0]
    if (wt.overwrite) wt<-n.u[n.u>0]
}
### More validation stuff
n <- length(x)
if (n <= 1) {
if (!full) return (y)
else return(list(x=x,y=y,z=x))
}
if (any(is.na(x)) || any(is.na(y))) {
    stop ("Missing values in 'x' or 'y' not allowed")    }
if (any(diff(x)<=0)) {stop ("x must be strictly increasing")}

z<-x  # Keeping a 'clean' copy of x for final stage
yy=y  # Keeping a 'clean' copy of y, as well
if (dec) y = -y

lvlsets <- (1:n)
#### Main iteration loop
repeat {

# Find adjacent violators
    viol <- (as.vector(diff(y)) <= 0)

    if (!(any(viol))) break
    i <- min( (1:(n-1))[viol]) # Pool first pair of violators
    y[i] <- (y[i]*wt[i]+y[i+1]*wt[i+1]) / (wt[i]+wt[i+1])
    x[i] <- (x[i]*wt[i]+x[i+1]*wt[i+1]) / (wt[i]+wt[i+1])  # new x is calculated
    wt[i]<-wt[i]+wt[i+1]  # weights are combined

# Deleting the i-1-th element
    y<-y[-(i+1)]
    x<-x[-(i+1)]
    wt<-wt[-(i+1)]
    n <- length(y)
    if (n <= 1) break  ### Cannot interpolate if only 1 point left
 }  # End iteration loop

if (boundary==1) {
# Utilize this option if you wish to use linear extrapolation outside the boundary
# (the 'approx' function does not have this option)
# In general, this is *not* recommended;  rather, impose boundary conditions whenever possible
# (as inputs or after output of this function)  or use the default, constant boundary conditions

    if (n==1) {
    x=c(xbounds[1],x,xbounds[2])
    y=c(ybounds[1],y,ybounds[2])
    wt=c(1,wt,1)
    n=3
} else {
    if (x[n]<max(z)) {
        x<-c(x,max(z))
        y<-c(y,y[n]+(y[n]-y[n-1])*(x[n+1]-x[n])/(x[n]-x[n-1])) }
    if (x[1]>min(z)) {
        x<-c(min(z),x)
        y<-c(y[1]-(y[2]-y[1])*(x[2]-x[1])/(x[3]-x[2]),y) }
        }
  } # End `boundary==1' case

# Now we re-interpolate to original x values, stored in z
# If we didn't set boundary=1, then this will give constant
# y values for x values falling outside new range of x

if (dec) y = -y

if (n==1) {
    if (!full) return(y)
    else return(list(output.y=rep(y,length(z)),original.x=z,original.y=yy,alg.x=x,alg.y=y,alg.wt=wt))
}
if (!full) return(approx(x,y,z,rule=boundary)$y)
else return(list(output.y=approx(x,y,z,rule=boundary)$y,original.x=z,original.y=yy,alg.x=x,alg.y=y,alg.wt=wt))
}
\end{verbatim}
\end{small}
\subsection*{Random-$F$ Simulation}
\begin{small}
\singlespacing
\begin{verbatim}

######## Function to generate random scenarios of a binary toxicity-response CDF F
######## Assaf P. Oron, Summer 2012
sceneMaster=function(nlev,marg=round(nlev/2),baseline=0.25,peakmean=3,peaksd=4,targ=.3,maxstep=2.5/nlev,minstep=.15/nlev,maxerr=.5/nlev,minedge=maxerr,protectfac=1.5,warp=TRUE,shift=0)
{
# nlev (integer): number of dose levels
# marg (integer): maximum extent of the "padding" margin that inflates the number of dose levels to choose our actual levels from
# baseline (positive continuous): minimum value of the Dirichlet parameter at any level
# peakmean (positive continuous): the approximate relative size of the jump in Dirichlet parameters around the density mode.
# peaksd (positive continuous): Quantifies the mode's sharpness.
# targ (continuous in (0,1)): the target toxicity rate. Used to calibrate the asymmetry of the dose set
# maxstep (continuous in (0,1)): the size of the largest allowed jump between adjacent doses.
# maxstep (continuous in (0,1)): the size of the smallest allowed jump between adjacent doses.
# maxerr (continuous in (0,1)): the size of the largest allowed difference between F(MTD) and 'targ'.
# minedge (continuous in (0,1)): the smallest advantage (in closeness to 'targ') that the MTD has over other levels.
# protectfac (positive continuous): also related to the MTD's advantage. No other level is allowed to be closer to 'targ' than 'protectfac*maxerr'.
# warp (logical): should the candidate set of F values be raised to a random power in (targ,1/targ) before final vetting?
# shift (continuous): the maximum amount of random constant shift that the F set receives before final vetting. If zero, this added randomization is skipped.

require(gtools)  # a library having the Dirichlet randomization function

accept=FALSE
while(!accept)   # Iterate until scenario passes vetting
{
	accept=TRUE
	peak=rlnorm(1,meanlog=log(peakmean),sdlog=log(peaksd)) # relative magnitude of mode (roughly speaking)
	seedlen=nlev+2*sample(1:marg,size=1) # We generate a Dirichlet vector longer than nlev
	asym=round(2*seedlen*(targ-0.5)) # asymmetry factor to help center the MTD distribution somewhat
	
# Dirichlet vector parameters: sum of a roughly-uniform base and a Gaussian peak
  	seeds=runif(seedlen,baseline,baseline*2)+2.5*peak*dnorm(1:seedlen,mean=sample(1:seedlen,size=1,prob=dnorm(1:seedlen,mean=(seedlen-asym)/2,sd=seedlen)),sd=peaksd*runif(1,seedlen/8,seedlen/2))

	#print(round(seeds,1))
# Now the actual Dirichlet vector, and immediately selecting nlev elements from it
# This sample too is asymmetric to improve eventual centering on target
	sampasym=sign(asym)*min(abs(asym)-1,seedlen-nlev-1)

	candid=cumsum(rdirichlet(1,seeds))[sort(sample(max(1,sampasym):min(seedlen-1,seedlen-1+sampasym),size=nlev))]
	
# Adding a random stretch!
	if(warp)	candid=candid^runif(1,.1/targ,1/targ)
# Adding a random shift!
	if (shift>0)	candid=candid+runif(1,max(-shift,-candid[1]),min(1-candid[nlev],shift))

#  cat ("After: ")
# print(round(candid,2))
# Checking acceptability
	gaps=diff(candid)	
# Rejecting if between-level increment too large/small anywhere
	if(max(gaps)>maxstep || min(gaps)<minstep) accept=FALSE

# Rejecting if best level too far from target.
  	closest=min(abs(candid-targ))
	if(closest>maxerr) accept=FALSE
# ... or has close competition
  	if(min(abs(candid-targ)[-which.min(abs(candid-targ))])<max(minedge+closest,protectfac*maxerr)) accept=FALSE
} 	
return(candid)
}
		
\end{verbatim}
\end{small}

\end{document}